\documentclass[journal,onecolumn, draftclsnofoot, 12pt]{IEEEtran}
\IEEEoverridecommandlockouts
\usepackage{lipsum} 
\usepackage{cite}
\usepackage{xspace}
\usepackage{subcaption}
\usepackage[cmex10]{amsmath}
\usepackage{mathtools,amssymb,lipsum}
\usepackage{lettrine}
\usepackage{bm}
\usepackage{algorithm,algorithmic}
\usepackage{array}
\usepackage{url}  
\usepackage{amsthm}
\usepackage{multirow}
\usepackage{color}
\usepackage{epstopdf}
\usepackage{endnotes}
\usepackage{framed}
\usepackage{cool}
\usepackage{enumerate}
\usepackage{url}
\usepackage[subnum]{cases}
\usepackage[nocomma]{optidef}
\usepackage{etoolbox}
\usepackage{amsmath,amssymb,amsfonts}
\usepackage{graphicx}
\usepackage{textcomp}
\usepackage{mathtools}
\usepackage[thinc]{esdiff}
\begin{document}

\title{Joint UAV Deployment and Resource Allocation in THz-Assisted MEC-Enabled Integrated Space-Air-Ground Networks} 

%{Shell \MakeLowercase{\textit{et al.}}: Bare Advanced Demo of IEEEtran.cls for IEEE Computer Society Journals} 

\author{Yan~Kyaw~Tun,~\IEEEmembership{Member,~IEEE},~György~Dán, ~\IEEEmembership{Senior~Member,~IEEE},~Yu~Min~Park, and~Choong~Seon~Hong,~\IEEEmembership{Fellow,~IEEE}
\thanks{Yan Kyaw Tun is with the Department of Electronic Systems, Aalborg University, A. C. Meyers Vænge 15, 2450 København, email: ykt@es.aau.dk.} 
\thanks{György Dán is with the Division of Network and Systems Engineering, School of Electrical Engineering and Computer Science, KTH Royal Institute of Technology, Brinellvägen 8, 114 28 Stockholm, Sweden, e-mail: gyuri@kth.se}.
\thanks{Yu Min Park, and Choong Seon Hong  are with the Department of Computer Science and Engineering, Kyung Hee University,  Yongin-si, Gyeonggi-do 17104, Rep. of Korea, e-mails:{\{yumin0906, cshong\}@khu.ac.kr}.}}

%{Shell \MakeLowercase{\textit{et al.}}: Bare Advanced Demo of IEEEtran.cls for IEEE Computer Society Journals}

\maketitle
\begin{abstract}
Multi-access edge computing (MEC)-enabled integrated space-air-ground (SAG) networks have drawn much attention recently, as they can provide communication and computing services to wireless devices in areas that lack terrestrial base stations (TBSs). Leveraging the ample bandwidth in the terahertz (THz) spectrum, in this paper, we propose MEC-enabled integrated SAG networks with collaboration among unmanned aerial vehicles (UAVs). We then formulate the problem of minimizing the energy consumption of devices and UAVs in the proposed MEC-enabled integrated SAG networks by optimizing tasks offloading decisions, THz sub-bands assignment, transmit power control, and UAVs deployment. The formulated problem is a mixed-integer nonlinear programming (MILP) problem with a non-convex structure, which is challenging to solve. We thus propose a block coordinate descent (BCD) approach to decompose the problem into four sub-problems: 1) device task offloading decision problem, 2) THz sub-band assignment and power control problem, 3) UAV deployment problem, and 4) UAV task offloading decision problem. We then propose to use a matching game, concave-convex procedure (CCP) method, successive convex approximation (SCA), and block successive upper-bound minimization (BSUM) approaches for solving the individual subproblems. Finally, extensive simulations are performed to demonstrate the effectiveness of our proposed algorithm.
\end{abstract}


\IEEEpeerreviewmaketitle
	
\section{Introduction}
\IEEEPARstart{I}{n}ternet of Things (IoT) devices are expected to be deployed worldwide for performing latency-sensitive tasks with significant computation requirements, such as autonomic navigation, road traffic monitoring, forest fire monitoring, and rescue operations in disaster areas~\cite{ren2017serving}. However, it is problematic for energy-constraint IoT devices to execute complex tasks on time locally. Edge computing could enable the devices to execute their tasks on time by offloading the tasks to computing servers deployed at terrestrial base stations (TBSs) and access points (APs), but terrestrial networks may not be available in remote areas and in disaster areas.

MEC-enabled integrated SAG networks have recently emerged as a potential technology for providing remote computation services to IoT devices in areas where there is no terrestrial infrastructure~\cite{liu2018space}\cite{dai2020dynamic}. Integrated SAG networks can leverage the computational and communication resources of unmanned aerial vehicles (UAVs) and of low earth orbit (LEO) satellites for providing pervasive access to computing services. 

%A potential advantage of SAG networks compared to TBS-based MEC is that UAVs can have a line-of-sight communication with IoT devices and hence can rely on the THz spectrum, which may enable latency-sensitive applications. THz communication is, however, most suitable for short range communication due to severe link attenuation and dispersion \cite{akyildiz2018combating}, hence the placement of UAVs and the allocation of communication resources (sub-bands and transmit power) to devices need to be optimized together with the allocation of computing tasks \cite{xu2021joint} \cite{park2022joint} \cite{9204738}. The coupling between task allocation and communication resource management makes it challenging to optimize SAG networks for latency-sensitive tasks. 

A key requirement for SAG to become successful is high bitrate connectivity between IoT devices and UAVs. A promising candidate for this purpose could be  THz communication, ranging from $0.1$ to $10$ THz, since it can provide higher bitrates due to the vast spectrum than what is achievable at lower frequency bands~\cite{akyildiz2022terahertz}\cite{chaccour2022seven} \cite{shafie2022terahertz}. The main detriment of relying on the THz band is severe link attenuation, which is combined with high dispersion \cite{akyildiz2018combating} and the easy obstruction of communication links through objects. Thus, the use of the THz frequency band in SAG networks could be feasible for short-range aerial communication, i.e., communication between UAVs and ground IoT devices to offload the devices' tasks to the servers attached to UAVs for further processing, due to the existence of line-of-sight (LoS) communication links in the Air-to-Ground communication \cite{yuan2020potential}. However, its efficient use requires joint consideration of UAV deployment for optimizing LoS communication links between UAVs and ground IoT devices, and the optimization of wireless resources, such as sub-band allocation and transmit power control~\cite{xu2021joint} \cite{park2022joint} \cite{9204738}. 

In this paper, we address the above challenge, considering energy efficient task offloading in MEC-enabled integrated SAG networks. The considered architecture adopts the THz frequency band for aerial base stations (i.e., UAVs) to provide remote wireless access to the ground wireless devices for offloading their computation tasks to the edge servers installed at UAVs. Importantly, the proposed architecture allows collaboration among UAVs, i.e., UAVs can decide whether to relay computational tasks among each other or to offload them to LEO satellites. As a result of collaboration among the UAVs, the energy consumption of the devices can be further reduced. 

To the best of our knowledge, this paper is the first to study the energy minimization problem in THz-assisted MEC-enabled integrated SAG networks,  incorporating UAV collaboration by concurrently optimizing task offloading decisions of UAVs and devices, THz sub-bands assignment, transmit power control, and UAV deployment. The main contributions of this paper are as follows: 
	
	\begin{itemize}
	    \item We first formulate the energy minimization problem in THz-assisted MEC-enabled integrated space-air-ground networks by optimizing the offloading decisions of the devices and the UAVs, THz sub-bands assignment and transmit power control, UAVs deployment, while satisfying the delay constraint of each device's task, resource constraints of the THz band, and the transmit power constraint of each device.
	    
	    \item Secondly, we show that the formulated problem is a non-convex mixed integer programming problem due to the coupling of decision variables in the objective function and constraints. To obtain a solution, we divide the problem into four sub-problems using the block coordinate descent (BCD) method: 1) device task offloading decision problem, 2) the THz sub-band assignment and power control problem, 3) UAV deployment problem, and 4) UAV task offloading decision problem.
	    
	    \item Thirdly, we show that the device task offloading decision problem is convex and then propose the standard optimization technique to solve the problem. Then, a one-to-one matching game and CCP approach are proposed to solve sub-bands assignment and power control problems. Finally, SCA and BSUM methods are proposed to solve the UAVs deployment and UAVs tasks offloading decision problems, respectively. 
	    
	    \item Finally, we demonstrate the convergence of the proposed algorithm by using extensive simulations. Furthermore, to show the effectiveness of our proposed algorithm, we compare the results of our proposed algorithm to the baseline schemes proposed in recent literature  \cite{zhu2020joint} and \cite{mao2020joint}.
	\end{itemize}
	
The rest of this paper is organized as follows. The related works and system model are described in Section \ref{rela} and Section \ref{sys}, respectively. Section \ref{prob} presents the problem formulation and the proposed solution is presented in Section \ref{sol}. Simulation results are shown in Section \ref{simu}. Section \ref{con} concludes the paper.

\section{Related Works}
\label{rela}
\subsection{Multi-Access Edge Computing (MEC)-Enabled Integrated Space-Air-Ground Networks}
MEC-enabled integrated SAG networks have received increasing attention in the recent literature \cite{chen2021energy, zhou2019delay, shi2022inter, wang2019radio, yu2021ec, chen2022energy, tun2022collaboration}. In \cite{chen2021energy}, the authors studied robust optimization-based UAV trajectory optimization and power control in SAG networks. Moreover, the work \cite{zhou2019delay} investigated linear programming-based UAV trajectory optimization and task offloading scheme. However, both \cite{chen2021energy} and  \cite{zhou2019delay} only took into account a single UAV scenario, leaving out power control, resource allocation, and interference management. In \cite{shi2022inter}, the authors proposed greedy and SCA-based task offloading and UAVs deployment schemes in the integrated SAG networks. The work \cite{wang2019radio} proposed radio resource allocation and task offloading framework for the integrated SAG vehicular networks. The authors in \cite{yu2021ec} investigated a machine learning-based framework for the MEC-enabled integrated SAG networks in order to offer computation services to numerous internet of vehicles (IoVs) in remote regions. In \cite{chen2022energy}, the authors studied SCA-based hybrid task offloading and computing resource allocation scheme in the SAG networks. However, power control, interference management, and collaboration among UAVs were omitted in \cite{ shi2022inter, wang2019radio, yu2021ec, chen2022energy }. The authors in \cite{tun2022collaboration} introduced the collaboration among UAVs in the multi-UAV-assisted MEC system. However, deployment of the UAVs, power control, interference management, and MEC-enabled satellites were omitted. 

All of the aforementioned works, however, made the assumption that their proposed SAG networks would operate in the sub-6 GHz frequency band. With the rapid growth of connected wireless devices and the limited available bandwidth (i.e., communication resource) at the sub-6 GHz band, the maximum bandwidth usage at the considered frequency band has been reached. Thus, researchers are eager to explore the untouched THz frequency band with available abundant bandwidth to fill the resource requirement of the devices in future wireless networks.

\subsection{THz-assisted Multi-Access Edge Computing}
The management of the THz spectrum for MEC was considered in \cite{mamaghani2022terahertz, du2020mec, liu2021learning, chaccour2022can, chaccour2019reliability, xie2020reliable,  chaccour2020ruin, busari2019generalized, pan2022self,   liu2022proximal, huang2021multi}. In \cite{mamaghani2022terahertz}, the authors proposed a secure mobile relaying system with UAV assistance that gathers data from several ground user equipment (UEs) and sends it to a destination using THz bands. The work \cite{du2020mec} presented a viewpoint rendering offloading decision and transmit power control technique based on deep reinforcement learning for virtual reality (VR) video streaming via THz-wireless channels. Furthermore, in \cite{liu2021learning}, the authors proposed a machine learning-based phase-shift design of IRS elements and rendering transmission for the VR system via IRS-assisted THz networks. The authors proved that using the THz frequency band could satisfy the ultra-reliable and low-latency requirement of the VR system in \cite{chaccour2022can}, and \cite{chaccour2019reliability}. The work \cite{xie2020reliable} discussed an optimization technique-based framework for the transmit power control and task offloading via THz frequency in the MEC system. In \cite{chaccour2020ruin}, the authors discussed the ruin theory-based age of information (AoI) minimization scheme in augmented reality (AR) system over THz networks. The work \cite{busari2019generalized} presented a hybrid beamforming scheme for the vehicular networks over THz massive MIMO system. In \cite{pan2022self}, the authors proposed a penalty-constrained convex approximation (PCCA)-based framework for transferring data and power concurrently over THz networks. In \cite{liu2022proximal}, distributed proximal policy optimization (DPPO) based beamforming and phase-shift design for the IRS-assisted cooperative communication and sensing system over THz networks were examined. The study \cite{huang2021multi} presented a multi-hop IRS-assisted THz communication system beamforming architecture based on deep reinforcement learning. 

All of the aforementioned existing works separately considered THz-assisted wireless networks and MEC-enabled integrated SAG networks. As a result, in contrast to previously published studies, we explore MEC-enabled integrated SAG networks over the THz frequency band in this paper. Additionally, we consider collaboration among UAVs in the proposed THz-assisted MEC-enabled integrated SAG networks, which has never been taken into account in all of the existing works.
	\section{System Model}
	\label{sys}
    \begin{figure}[t!]
    \centering
    \captionsetup{justification = centering}
    \includegraphics[width=0.5\linewidth]{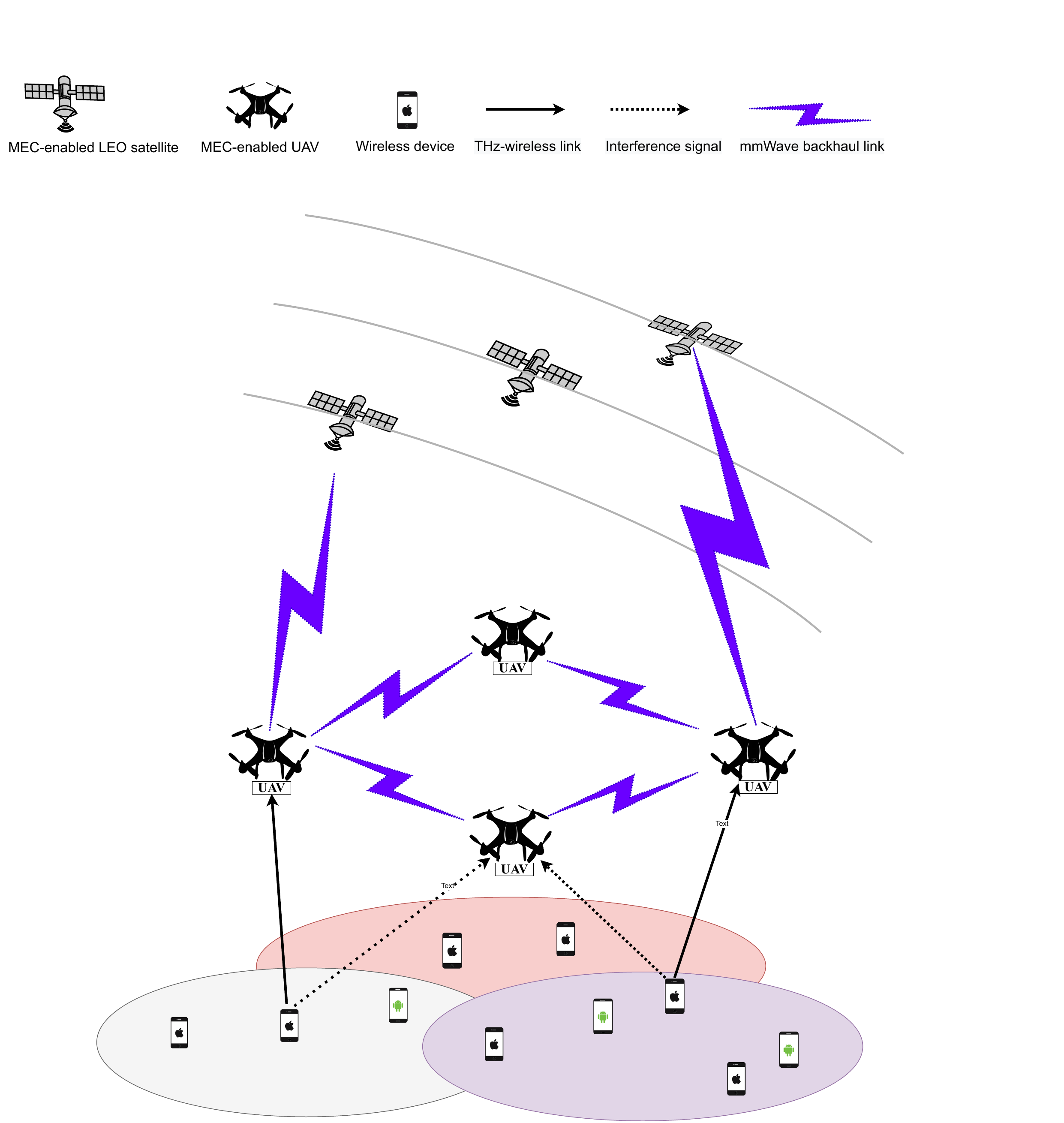}  %{Untitled Diagram.drawio (6).pdf}
    \caption{Illustration of MEC-enabled integrated space-air-ground networks.}
    \label{sysmod}
    \end{figure}
	
We consider a MEC system in the integrated space-air-ground network that consists of a set $\mathcal{J}$ of $J$ wireless devices, a set $\mathcal{K}$ of $K$ UAVs, and a set $\mathcal{S}$ of $S$ LEO satellites, as illustrated in Fig.~\ref{sysmod}. Each device $ j \in \mathcal{J}$ in the considered network has a latency-sensitive computation task $T_j$, which can be characterized by a tuple $T_j = \{\varphi_j, \alpha_j, A_j\}$, where $\varphi_j$ is the maximum tolerable delay of the task, $\alpha_j$ is the CPU cycles needed to compute one bit of data, and $A_j$ is the data size of the task. Devices are energy constrained, we thus consider that each device offloads a certain amount of data of its computation task to its associated UAVs. In this paper, we assume that the association between devices and UAVs is already determined depending on the distance via the K-means algorithm.

We use $\mathcal{J}_k$ to denote the set of devices that offload a certain amount of data of their computation tasks to UAV $k$, and assume $ \mathcal{J} = \bigcup\limits_{k=1}^{K} \mathcal{J}_{k} $ where $\mathcal{J}_{k} \cap \mathcal{J}_{k'} = \emptyset, \forall k,k'\in \mathcal{K}, k \neq k'$. We consider that the THz frequency band is adopted for communication between devices and UAVs due to the abundance of bandwidth in this frequency band. The available bandwidth in the considered THz frequency band is divided into a set $\mathcal{B} $ of $B$ sub-bands, and we use $\omega$ to denote the bandwidth of each sub-band. Since, the THz frequency band is only suitable for short-range communication due to severe link attenuation and dispersion, we consider that mmWave (28 GHz) backhaul links are adopted to communicate among UAVs and satellites. Furthermore, we consider that UAVs and satellites can obtain the channel state information (CSI) of associated devices and UAVs,
e.g., using techniques presented in \cite{fan2019channel, guo2020prediction, chang2022csi}.

\subsection{Local Computing Model}   
Let $(A_j - \beta_j^k)$ be the amount of data of device $j$'s task that is processed locally on device $j$ and $\beta_j^k$ be the amount of data that is offloaded to the associated UAV $k\in \mathcal{K}$ for remote computing. Thus, wireless device $j$'s local computation delay for completing the task which is calculated by \cite{9204738} as
\begin{equation}
   l_j^ {k, \textrm{loc}} = \frac{(A_j - \beta_j^k)\alpha_j}{f_j}, \forall j \in \mathcal{J},
\end{equation}
where $f_j$ represents the computation capacity of device $j$. The local energy usage of wireless device $j$ which is expressed in \cite{9204738} as
\begin{equation}
    E_j^{k, \textrm{loc}} =  \kappa_j (f_j)^2\alpha_j(A_j - \beta_j^k), \forall j \in \mathcal{J}, 
\end{equation}
where $\kappa_j$ is a constant that depends on the wireless device's chip architecture.
	
\subsection{Communication Model}
    Each device uses one of the available THz sub-bands at its associated UAV for data transmission for offloading. We define $a_j^{k,b} \in \{ 0,1\}$ as the sub-band assignment variable, which represents whether or not sub-band $b$ is assigned to device $j$ associated to UAV $k$, i.e.,  
	\begin{equation}
	a_j^{k,b} =
	\begin{cases}
	1, \ \ \text{if sub-band $b$ is assigned to device $j$, which} \\ 
      \ \ \ \text{offloads $\beta_j^k$ amount of data of its task to UAV $k$}, \\
	0, \ \ \text{otherwise}.
	\end{cases}
	\end{equation}
      We consider that the orthogonal frequency division multiple access (OFDMA) scheme is used for communication between a UAV and its associated devices in order to avoid intra-cell interference. At the same time, to improve spectrum efficiency, we consider frequency reuse between UAVs, i.e., all UAVs operate on the same frequency to communicate with their associated devices. Thus, inter-cell interference between different UAVs may exist. As a result, in each UAV, a sub-band can be assigned to at most one device,
	\begin{equation}
	 \sum\limits_{j \in \mathcal{J}_k} 	a_j^{k,b} \leq 1, \forall b \in \mathcal{B}, \forall k \in \mathcal{K}. 
	\end{equation}
Furthermore, we assume that at most one sub-band can be assigned to a device,
	\begin{equation}
	 \sum\limits_{b \in \mathcal{B}} 	a_j^{k,b} \leq 1, \forall j \in \mathcal{J}_k, \forall k \in \mathcal{K}.   
	\end{equation}
	
We can express the channel gain of device $j$ on sub-band $b$ to UAV $k$ as \cite{xu2021joint}
\begin{equation}
    g_j^{k,b} = g_0(d_j^k)^{-2} e^{-i_b(f)d_j^k}, \forall j \in \mathcal{J}_k, \forall b \in \mathcal{B}, \forall k \in \mathcal{K}, 
\end{equation}
where $g_0$ is the channel gain at reference distance $d=1$ m, $i_b(f)$ is the coefficient of molecular absorption, which is influenced by both the concentration of water vapor molecules in the air and the network operating frequency (i.e., THz frequency), and $d_j^k$ is the distance between device $j$ and UAV $k$, which can be computed as
\begin{equation}
	d_j^k = \sqrt{ (x_k - x_j)^2 + (y_k - y_j)^2 + h_k^2}, \forall j \in \mathcal{J}_k, \forall k \in \mathcal{K},
\end{equation}
where $\boldsymbol{\ddot{c}}_j= [x_j, y_j]^T$ and $\boldsymbol{o}_k = [x_k, y_k]^T$ are the horizontal coordinates of device $j \in \mathcal{J}$ and UAV $k \in \mathcal{K}$ respectively, and $h_k$ is the hovering altitude of the UAV.

The received signal to interference plus noise ratio (SINR) between device $j$ on sub-band $b$ and its associated UAV $k$ is then given by \cite{xu2021joint} as
\begin{equation}
	\gamma_j^{k,b} = \frac{P_j^{k,b} g_j^{k,b}}{ I_j^{k,b}+ \sigma^2},  \forall  j \in \mathcal{J}_k, \forall b \in \mathcal{B}, \forall k \in \mathcal{K}, 
\end{equation}
where $P_j^{k,b}$ represents the transmit power of device $j$, $\sigma^2$ is the additive white Gaussian noise power, and 
\begin{equation}
     I_j^{k,b} = \sum\limits_{k' \in \mathcal{K}, k' \neq k} \sum\limits_{j' \in \mathcal{J}, j' \neq j} P_{j'}^{k',b} g_{j'}^{k,b}
\end{equation}
is the interference experienced at UAV $k$. Finally, we can calculate the achievable data rate of device $j$ on sub-band $b$ as 
\begin{equation}
    R_j^{k,b} = \omega \log_2(1+ \gamma_j^{k,b}), \forall j \in \mathcal{J}_k, \forall b \in \mathcal{B}, \forall k \in \mathcal{K}. 
\end{equation}
We use (10) to compute the data rate $R_j^k = \sum\limits_{b \in \mathcal{B}} a_j^{k,b}  R_j^{k,b}$ of device $j$, which can be used for computing the transmission delay experienced by device $j$ when offloading $\beta_j^k$ amount of data of its task to UAV $k$  which is described by \cite{9204738} as
\begin{equation}
    l_j^{k, \textrm{trans}} = \frac{\beta_j^k}{R_j^k}, \forall j \in \mathcal{J}_k, \forall k \in \mathcal{K}.
\end{equation}
We express the transmission energy consumed at device $j$ when offloading a $\beta_j^k$ amount of data of its task to UAV $k$ as \cite{9204738}
\begin{equation}
	E_j^{k, \textrm{trans}} = l_j^{k, \textrm{trans}} \sum_{b \in \mathcal{B}} P_j^{k,b},  \forall j \in \mathcal{J}_k, \forall k \in \mathcal{K}.
\end{equation}
After receiving the offloaded data from its associated devices, UAV $k$ decides to either process them locally on its server or transfer them to neighboring UAVs or to the satellites.

Thus, we introduce the binary decision variable $x_j^{k \rightarrow k'} \in \{0,1\}$, indicating whether or not the offloaded data of device $j$ is transferred to neighboring UAV $k' \in \mathcal{K}$,
	\begin{equation}
	v_{j}^{k \rightarrow k'} = 
	\begin{cases}
	1, \ \ \text{ if offloaded data of device $j$ is transferred } \\
         \ \ \ \text{from UAV $k$ to UAV $k'$}, \\ 
	0, \ \ \text{otherwise}.
	\end{cases} \\
	\end{equation}
	We use $\beta^{k \rightarrow k'} = {\sum\limits_{j\in \mathcal{J}_k} v_j^{k \rightarrow k'} \beta_j^k}$ to denote the total amount of data transferred from UAV $k$ to UAV $k'$. The transmission time from UAV $k$ to $k'$ is determined by the achievable channel gain between these UAVs and can be expressed as \cite{tun2022collaboration}
	\begin{equation}
	\Gamma^{k \rightarrow k'} = \frac{P^{k \rightarrow k'} g_k^{\textrm{tx}} g_{k'}^{\textrm{rx}}L_r}{t_nHB_{\textrm{mm}}^{k \rightarrow k'}}   \left( \frac{c}{4\pi d_k^{k'} f_c^{\textrm{mm}}}\right)^2, 
	\end{equation}
	where $P^{k\rightarrow k'}$ is the transmit power of UAV $k$, $g_k^{\textrm{tx}}$ and $g_{k'}^{\textrm{rx}}$ represent the antenna gains of the transmitter, UAV $k$, and the receiver UAV $k'$, $L_r$ is the amplification factor, $t_n$ is the noise temperature, $H$ is Boltzmann's constant, $f_c^{\textrm{mm}}$ is the mmWave carrier frequency, $ B_{\textrm{mm}}^{k\rightarrow k'}$ is the available bandwidth between UAV $k$ and UAV $k'$, and $d_k^{k'}$ denotes the distance between UAV $k$ and $k'$. Then, the achievable backhaul capacity between UAV $k$ and $k'$ is given by
	\begin{equation}
	 R^{k\rightarrow k'} = B_{\textrm{mm}}^{k\rightarrow k'} \log_2\left(1 + \Gamma^{k\rightarrow k'}\right), \forall k, k' \in \mathcal{K}.
    \end{equation}
	Finally, the transmission delay assuming that device j's data is transferred to UAV k' is as \cite{tun2022collaboration}
	\begin{equation}
	l_j^{k \rightarrow k',\textrm{trans}} = \frac{\beta^{k \rightarrow k'}}{R^{k \rightarrow k'}}, \forall k, k' \in \mathcal{K}.
	\end{equation}
   Moreover, the total transmission energy consumed at UAV $k$ when transferring the data of its associated devices to nearby UAV $k'$ is given by \cite{tun2022collaboration} as
	\begin{equation}
	E^{k \rightarrow k',\textrm{trans}} =P^{k \rightarrow k'} \left(\frac{\beta^{k \rightarrow k'}}{R^{k \rightarrow k'}}\right),  \forall k,k' \in \mathcal{K}.
	\end{equation} 

 Finally, we introduce the binary decision variable $z_j^{k \rightarrow s} \in \{0,1\}$, to indicate whether or not UAV $k$ transfers the offloaded data of device $j$ to satellite $s$
	\begin{equation}
	z_{j}^{k \rightarrow s} = 
	\begin{cases}
	1, \ \ \text{if offloaded data of device $j$ is transferred} \\
          \  \  \text{from UAV $k$ to satellite $s$,}\\
	0, \ \ \text{otherwise}.
	\end{cases}
	\end{equation}
   Let $\beta^{k \rightarrow s} = \sum\limits_{j \in \mathcal{J}_k} z_{j}^{k \rightarrow s} \beta_j^k$ be the total amount of data transferred from UAV $k$ to satellite $s$. Then, the transmission delay incurred when device $j$'s offloaded data is transferred from UAV $k$ to satellite $s$ can be expressed as \cite{tun2023collaborative}
	\begin{equation}
	l_j^{k \rightarrow s, \textrm{trans}} = \frac{\beta^{k \rightarrow s}}{R^{k \rightarrow s}}, \forall k \in \mathcal{K}, \forall s \in \mathcal{S},
	\end{equation}
   where $R^{k \rightarrow s}$ is the achievable backhaul link capacity between the UAV and the satellite that can be calculated based on (15). Additionally, the amount of transmission energy used by UAV $k$ when transferring the total offloaded data of its associated devices to the satellite $s$ is given by \cite{tun2023collaborative}
	\begin{equation}
	E^{k \rightarrow s, \textrm{trans}} =P^{k \rightarrow s} \left(\frac{\beta^{k \rightarrow s} }{R^{k \rightarrow s}}\right),  \forall k \in \mathcal{K}, \forall s \in \mathcal{S}.
	\end{equation}

\subsection{Remote Computing Model}

In order to express whether or not the offloaded data of device $j$ is computed at UAV $k$, we define the binary decision variable, $w_j^{j \rightarrow k} \in \{0,1\}$, i.e., 
	\begin{equation}
	w_{j}^{k} = 
	\begin{cases}
	1, \ \ \text{ if offloaded data of device $j$ is computed at UAV $k$}, \\
	0, \ \ \text{otherwise}.
	\end{cases}
	\end{equation}
If indeed the wireless device $j$'s offloaded data is computed at UAV $k$, i.e., $w_{j}^{k}=1$, then the computation delay is \cite{9204738}
	\begin{equation}
	l_j^{k, \textrm{comp}} = \frac{\alpha_j\beta_j^k}{f_j^k}, \forall j \in \mathcal{J}_k, \forall k \in \mathcal{K},
	\end{equation}
	where $f_j^k$ is the computation capacity of UAV $k$ that is allotted to compute the offloaded data of wireless device $j$. We consider that the UAVs use proportional allocation \cite{tun2019wireless} as
	\begin{equation}
	f_j^k =  \frac{\alpha_j\beta_j^k}{\sum\limits_{j' \in \mathcal{J}} w_{j'}^{k} \alpha_{j'}\beta_{j'}^k} F_k^{\mathbf{max}},  
	\end{equation}
   where $F_k^{\mathbf{max}}$ denotes the computation capacity of UAV $k$. As a result, when wireless device $j$ offloads $\beta_j^k$ amount of data of its computation task to UAV $k$, the total delay it encounters is
	\begin{equation}
	l_j^{k, \textrm{remote}} = l_j^{k, \textrm{trans}}  + l_j^{k, \textrm{comp}}, \forall j \in \mathcal{J}_k, k \in \mathcal{K}.
	\end{equation}
	The energy usage at UAV $k$ for processing the offloaded data of wireless device $j$ can be written as \cite{9204738}
	\begin{equation}
	E_j^{k, \textrm{comp}} = \kappa (f_j^k)^2 \alpha_j\beta_j^k,  \forall j \in \mathcal{J}_k, \forall k \in \mathcal{K}, 
	\end{equation}
   where $\kappa$ is a constant that depends on the chip architecture of the UAV's MEC server. Consequently, the total delay that the device $j$ experiences when the offloaded data of its task is performed at UAV $k'$ is as
	\begin{equation}
	\begin{split}
	l_j^{k \rightarrow k',\textrm{remote}} &= l_j^{k, \textrm{trans} } + l_j^{k \rightarrow k',\textrm{trans}} + l_j^{k \rightarrow k', \textrm{comp}}, \forall j \in \mathcal{J}_k, \\
   & \ \ \ \ \ \ \ \ \ \  \ \ \ \ \ \ \ \ \  \forall k,k' \in \mathcal{K} , k \neq k'.
	\end{split}
	\end{equation}
  Finally, let $l_j^{k \rightarrow s, \textrm{comp}}$ denote the computation delay when wireless device $j$'s offloaded data is processed at satellite $s$ which can be calculated based on (22). Then, the total delay that wireless device $j$ encounters when its offloaded data is transferred to the satellite is
	\begin{equation}
	\begin{split}
	l_j^{k \rightarrow s,\textrm{remote}} &= l_j^{k, \textrm{trans}} + l_j^{k \rightarrow s, \textrm{trans}} + l_j^{k \rightarrow s, \textrm{comp}} + 2l_j^{k \rightarrow s, \textrm{pro}}, \\
  &\ \ \ \ \ \ \ \ \ \ \ \forall j \in \mathcal{J}_k, \forall k \in \mathcal{K}, \forall s \in \mathcal {S}, 
	\end{split}
	\end{equation}
where $ 2l_j^{k \rightarrow s, \textrm{pro}} =\frac{2d_k^s}{c}$ is the round-trip propagation delay between UAV $k$ and the satellite $s$. In this paper, we consider that the satellite has a renewable energy source. Thus, we disregard the satellite's energy usage for computing the data transferred from all UAVs. 

We make the reasonable assumption that the available computation capacity at the satellites is significantly greater than that at the UAVs and devices, and thus the computation time at the satellites is negligible compared to the computation time at the UAVs and devices. As a result, in our work, we do not account for the computation time of the satellites. Moreover, in the considered application scenario, the size of the output data after the offloaded task of each device has been executed at the MEC servers of UAVs and satellites is much less than the input data size of the offloaded task. We thus do not account for the downlink transmission time in the problem formulation. At the same time, our model accounts for the downlink propagation delay, which may be significant for satellite communication. Therefore, the downlink communication from the UAVs and satellites to the ground devices is disregarded in this study. Thus, the total delay encountered  by device $j$ when $\beta_j^k$ amount of data of its computation task is offloaded to the associated UAV $k$ for remote computing is as 
\begin{equation}
    l_j^{\textrm{k, Remote}} =w_j^{k} l_j^{k,\textrm{remote}} + \sum\limits_{ k'\in \mathcal{K}, k'\neq k} v_j^{k \rightarrow k'} l_j^{k \rightarrow k',\textrm{remote}} + \sum\limits_{ s \in \mathcal{S}} z_j^{k\rightarrow s} l_j^{k \rightarrow s, \textrm{remote}}, \forall j\in \mathcal{J}_k, \forall k \in \mathcal{K}. 
\end{equation}

Finally, the total amount of energy used by the UAV to execute the offloaded data of the  devices in the considered integrated SAG network is provided by
\begin{equation}
    E_k^{\textrm{Tot}} = \sum\limits_{ j \in \mathcal{J}} w_j^{k}  E_j^{k, \textrm{comp}} + \sum\limits_{ k'\in \mathcal{K}, k' \neq k}  E^{k \rightarrow k',\textrm{trans}} +  \sum\limits_{ s\in \mathcal{S}}  E^{k \rightarrow s,\textrm{trans}},  \forall k \in \mathcal{K}. 
\end{equation}

%where $E^{k, \text{hov}}$ is the hovering energy and is given by \cite{monwar2018optimized}
%\begin{equation}
%    E^{k,\textrm{hov}} = P^{k, \textrm{hov}} l^{k, \textrm{hov}}, \forall k \in \mathcal{K},
%\end{equation}
%where $l^{k, \textrm{hov}} = \max\limits_{j \in \mathcal{J}}(\varphi_j)$ is the hovering time of UAV $k$ and %$P^{k, \textrm{hov}}$ indicates the power consumed at UAV $k$ to hover at a fixed location, which is defined as
%\begin{equation}
%   P^{k,\textrm{hov}} = \frac{\iota \sqrt{\iota}}{\zeta_k\sqrt{0.5\pi\phi r^2\Lambda}}  l^{k,\textrm{hov}}, \forall k \in \mathcal{K},
%\end{equation}
%Here, $\phi$ denotes the number of rotors in UAV $k$, $r$ presents the diameter of each rotor, $\iota$ is the trust that scales with the mass of the UAV, $\zeta_k$ is the power efficiency of UAV , and $\Lambda$ stands for the density of air.
\section{Problem Formulation}
\label{prob}
Our objective is to jointly optimize the deployment of UAVs, the task offloading decision for the devices and the UAVs, the transmit power, and the assignment of communication resources with the aim of minimizing the energy consumption of the UAVs and the devices subject to the available wireless resources (i.e., sub-bands and transmit power) and computing time constraints. Thus, we define the objective function as 
\begin{equation}
\mathbf{Q}(\boldsymbol{o}, \boldsymbol{\beta}, \boldsymbol{P},  \boldsymbol{a}, \boldsymbol{w}, \boldsymbol{v}, \boldsymbol{z}) = \sum\limits_{ k \in \mathcal{K}} \sum\limits_{j \in \mathcal{J}} \left(E_j^{k, \textrm{loc}} + E_j^{k, \textrm{trans}}\right) + \sum\limits_{ k \in \mathcal{K}}E_k^{\textrm{Tot}}. 
\end{equation}
We can then formulate the proposed optimization problem as 

	\begin{mini!}[2]                 % maxi! = maximize
		{\boldsymbol{o, \beta, P, a, w, v, z}}                               % optimization variable
		{\mathbf{Q}(\boldsymbol{o, \beta, P, a, w, v, z})}{\label{opt:P1}}{\textbf{P:}} 
		\addConstraint{  l_j^{k, \textrm{loc}}  \leq \varphi_j, \forall j \in \mathcal{J}_k, \forall k \in \mathcal{K},}
		\addConstraint{  l_j^{k, \textrm{Remote}}  \leq \varphi_j, \forall j \in \mathcal{J}_k, \forall k \in \mathcal{K},}
		\addConstraint{ 0 \leq \beta_j^ k \leq A_j , \ \ \forall j \in \mathcal{J}, \forall k \in \mathcal{K},}
		\addConstraint{\sum_{b \in \mathcal{B}} a_j^{k,b} \leq 1 \ \ \forall j \in \mathcal{J}_k, \forall k \in \mathcal{K},} 
		\addConstraint{\sum_{j \in \mathcal{J}_k} a_j^{k,b} \leq 1 \ \ \forall b \in \mathcal{B}, \forall k \in \mathcal{K},}
		\addConstraint{ 0 \leq P_j^{k,b} \leq P_j^{\mathbf{max}}, \forall j \in \mathcal{J}_k, \forall k \in \mathcal{K},}
		\addConstraint{w_j^{k} + \sum_{\substack{k'\in \mathcal{K}, \\ k'\neq k}} v_j^{k \rightarrow k'} + \sum_{s \in \mathcal{S}} z_j^{k\rightarrow s} = 1, } \nonumber 
        \addConstraint{ \ \ \ \ \ \ \ \ \ \ \ \ \ \ \ \ \ \ \ \ \ \ \ \ \ \  \ \ \ \forall j \in \mathcal{J}_k,}
		\addConstraint{ a_j^{k,b} \in \{0,1\},\forall j \in \mathcal{J}_k, \forall b \in \mathcal{B},  \forall k \in \mathcal{K},}
		\addConstraint{w_j^{ k} \in \{0,1\}, \forall j \in \mathcal{J}_k, \forall k \in \mathcal{K},}
		\addConstraint{v_j^{k \rightarrow k'} \in \{0,1\}, \forall j \in \mathcal{J}_k, \forall k,k' \in \mathcal{K},}
		\addConstraint{z_j^{k \rightarrow s} \in \{0,1\}, \forall j \in \mathcal{J}_k, \forall k \in \mathcal{K}, \forall s \in \mathcal{S},}
		\addConstraint{X_k^{\mathbf{min}} \leq x_k \leq  X_k^{\mathbf{max}}, \forall k \in \mathcal{K}, }
		\addConstraint{Y_k^{\mathbf{min}} \leq y_k \leq  Y_k^{\mathbf{max}}, \forall k \in \mathcal{K}, }
  \vspace{-0.1in}
	\end{mini!}
where constraints (31b) and (31c) guarantee that a task is executed within the task's maximum tolerable delay, then constraint (31d) assures that the data size of the task that is offloaded to the associated UAV $k \in \mathcal{K} $ must be less than the total input data size of the task of device $j$. Constraints (31e) and (31f) ensure that each THz sub-band in a UAV can only be assigned to one device, and the same is true for each device associated with a UAV. Constraint in (31g) guarantees that the device's transmit power is less than its maximum available power. Constraint (31h) ensures that the offloaded data of the device is computed at a single location (i.e., at the associated UAV (or) one of the nearby UAVs (or) one of the LEO satellites). Moreover, (31i), (31j), (31k), and (31l) are the binary decision variables. Finally, limitations on the coordinates of each UAV are ensured by constraints (31m) and (31n).

\section{Solution Approach}
\label{sol}
Convex optimization techniques cannot be employed directly to address the optimization problem in (31) because decision variables are coupled in the objective function and in the constraints, the problem has nonlinear constraints and binary variables, and has a non-convex structure. We thus propose to use the block coordinate descent (BCD) approach to decompose the problem into four sub-problems: \emph{ 1) device task offloading decision problem, 2) sub-band assignment and transmit power control problem, 3) UAV deployment problem, and 4) UAV task offloading problem.} Then, the decomposed sub-problems are solved alternatingly.

\subsection{Device Task Offloading Decision}

For a given sub-band assignment and transmit power decision $\{\boldsymbol{P, a}\}$, deployment $\{\boldsymbol{o}\}$ of UAVs, and offloading decision $\{\boldsymbol{w, v, z}\}$ of UAVs, we can formulate the device task offloading decision problem as
	\begin{mini!}[2]                 % maxi! = maximize
		{\boldsymbol{\beta}}                               % optimization variable
		{\mathbf{Q}(\boldsymbol{\beta})}{\label{opt:P1}}{\textbf{P1:}} 
	     \addConstraint{  l_j^{k, \textrm{loc}}  \leq \varphi_j, \forall j \in \mathcal{J}_k, \forall k \in \mathcal{K},}
		\addConstraint{  l_j^{k, \textrm{Remote}}  \leq \varphi_j, \forall j \in \mathcal{J}_k, \forall k \in \mathcal{K},}
		\addConstraint{ 0 \leq \beta_j^ k \leq A_j , \ \ \forall j \in \mathcal{J}, \forall k \in \mathcal{K},}
	\end{mini!}
From problem \textbf{P1}, we can see that the objective function (32a) and the constraints (32b)-(32d) are linear. Thus, we can conclude that problem \textbf{P1} is convex. As a result, we can solve the problem using convex optimization techniques.

\subsection{Sub-band Assignment and Power Control}

For a given task offloading decision $\{\boldsymbol{\beta}\}$ of devices, a deployment $\{\boldsymbol{o}\}$ of UAVs, and offloading decision $\{\boldsymbol{w, v, z}\}$ of UAVs, we can formulate the sub-band assignment and power control problem as
	\begin{mini!}[2]                 % maxi! = maximize
		{\boldsymbol{P, a}}                               % optimization variable
		{\mathbf{Q}(\boldsymbol{P, a})}{\label{opt:P1}}{\textbf{P2:}} 
		\addConstraint{ \textrm{(31c), (31e)-(31g), (31i)}},
  \vspace{-0.1in}
	\end{mini!}
Unfortunately, the decision variables in \textbf{P2} are coupled in the objective function and in the constraints, and the problem has a combination of binary and continuous variables. Thus, problem \textbf{P2} is a mixed-integer nonlinear programming (MINLP) problem that is NP-hard. Therefore, we develop a polynomial time two-stage distributed approach to address P2, by combining a matching game to assign sub-bands, and the concave-convex procedure (CCP) approach to evaluate the power control at each UAV.

\textbf{Stage 1 (Sub-band Assignment)}. We want to maximize the total transmission rate of the devices because by doing so, we can decrease their transmission delay, i.e., constraint (31c), and transmission energy, i.e., the objective function, as indicated in (11) and (12), respectively. In other words, transmission energy and delay have an inverse relationship with data rate. As a result, we can formulate the sub-band assignment problem as a data rate (i.e., transmission rate between devices and UAVs) maximization problem. However, the sub-band assignment problem is a combinatorial integer programming problem. Thus, deploying centralized optimization techniques can cause significant overhead and complexity. As a result, we propose a low complexity distributed matching algorithm \cite{gu2015matching} to solve the problem. Since a wireless device can only have one sub-band assigned to it and a sub-band can only be assigned to a maximum of one device, we can model our sub-band assignment problem as a one-to-one matching game. We first provide the definition of the one-to-one matching game for sub-band assignment at each UAV $k \in \mathcal{K}$.

\textit{Definition 1: Given two disjoint sets of players, $\mathcal{J}_k$ and $\mathcal{B}$, the one-to-one matching game $\vartheta_k: \mathcal{J}_k \rightarrow \mathcal{B}$ for the sub-band assignment is defined as:} 

\textit{1) $\vartheta_k(b) \subseteq \mathcal{J}_k $ and $|\vartheta_k(b)| \in \{0,1\}$, \ $\forall b \in \mathcal{B}$;}

\textit{2) $\vartheta_k(j) \subseteq \mathcal{B} $ and $|\vartheta_k(j)| \in \{0,1\}$ , \ $\forall j \in \mathcal{J}_k$;} 

\textit{3) $j = \vartheta_k(b) \leftrightarrow b = \vartheta_k(j)$, $\forall b \in \mathcal{B}, \forall j \in \mathcal{J}_k$.} 

Here, $|\vartheta_k(.)|$ is a representation of the cardinality of the matching outcome $\vartheta_k(.)$. Conditions (1) and (2) in the definition ensure that a sub-band can only be assigned to one device at a time and that a device can only have one sub-band assigned to it. Furthermore, according to condition (3), if device $j$ is matched with sub-band $b$, then sub-band $b$ must also be matched with device $j$. The outcome of the one-to-one matching game is the assignment mapping between a set of devices $\mathcal{J}_k$ and sub-bands $\mathcal{B}$. 

Firstly, we define the preference function of device $j \in \mathcal{J}_k$ for sub-band $b \in \mathcal{B}$ and the preference function of sub-band $b \in \mathcal{B}$ for device $j\in \mathcal{J}_k$ as $\theta_j(b)$ and $\theta_b(j)$, respectively. The notation $ b_1 \succ_{j} b_2$ implies that device $j$ prefers sub-band $b_1$ over $b_2$, i.e., $\theta_j(b_1) >\theta_j(b_2),$ and at the same time the notation $ j_1 \succ_{b} j_2$ indicates that the sub-band prefers device $j_1$ over $j_2$, i.e., $\theta_b(j_1) >\theta_b(j_2)$.
\begin{algorithm}[t!]
	\caption{\strut One-to-One Matching Game-Based Sub-band Assignment Algorithm}
	\label{alg:profit}
	\begin{algorithmic}[1]
	    \STATE{\textbf{Input:} $\mathcal{J}_k$, $\mathcal{B}$; }
		\STATE{\textbf{Initialization:} Set $\mathcal{J}_k^{\textrm{un}} = \mathcal{J}_k $, $\mathcal{B}_j = \mathcal{B}, \forall j \in \mathcal{J}_k$, a set of devices requested to sub-band $b, \mathcal{J}_k^{b,\textrm{req}} = \emptyset$, and a set of rejected devices from sub-band $b, \mathcal{J}_k^{b,\textrm{rej}}, \forall b\in \mathcal{B}$;}
		\STATE{Construct the preference list of devices in $\mathcal{J}_k$ according to (34) by equally allocating its available transmit power to all sub-bands, i.e., $P_j^{k,b} = \frac{P_j^{\mathbf{max}}}{B}, \forall j \in \mathcal{J}_k$;}
		\STATE{Find a stable matching $\vartheta^{*}_k$}
		\STATE{\textbf{while} $\sum\limits_{b \in \mathcal{B}}\sum\limits_{j\in \mathcal{J}_k} q_{jb} \neq 0 $ \textbf{do}}
		\STATE{ \ \  \  \ \  \textbf{for} $ j=1$ to $|\mathcal{J}_k^{\textrm{un}}|$ \textbf{do}}
		
		\STATE{ \ \  \  \  \  \ \  \  \  \    Find $b = \underset{ b \in \mathcal{B} }{\mathrm{argmax}} \ \theta_j(b)  $.}
		\STATE{ \ \  \  \  \  \ \  \  \  \ Make a request to the UAV $k$ by setting $q_{jb}=1$.}
		\STATE{\ \  \  \ \ \textbf{end for}}
		\STATE{\ \  \  \ \ \textbf{for} $b=1$ to $B$ \textbf{do} }
		\STATE{ \ \  \  \  \  \ \  \  \  \ Update $\mathcal{J}_k^{b,\textrm{req}} \leftarrow \{ j: q_{jb}=1, \forall j \in \mathcal{J}_k\}$.}
		\STATE{ \ \  \  \  \  \ \  \  \  \ Construct the preference list of UAV for its} 
		\STATE{\ \  \  \  \  \ \  \  \  \ available sub-bands according to (35).}
		\STATE{\ \  \  \  \  \ \  \  \  \  Find $j = \underset{ j \in \mathcal{J}_k }{\mathrm{argmax}} \ \theta_b(j)$.}
		\STATE{\ \  \  \  \  \ \  \  \  \ Assign sub-band $b$ to device $j$.}
		\STATE{\ \  \  \  \  \ \  \  \  \ Update $\mathcal{J}_k^{b,\textrm{rej}} \leftarrow \{\mathcal{J}_k^{b,\textrm{req}}\setminus j\}$. }
		\STATE{\ \  \  \  \  \ \  \  \  \ Update $\mathcal{B}_j \leftarrow \{\mathcal{B}_j\setminus b\}$, $\forall j \in \mathcal{J}_k^{b,\textrm{rej}} $.}
		\STATE{\ \  \  \ \ \textbf{end for}}
		\STATE{\ \  \  \ \ Update $\mathcal{J}_k^{\textrm{un}} \leftarrow \mathcal{J}_k^{\textrm{un}} \cap \{ \mathcal{J}_k^{1,\textrm{rej}} \cup ..... \cup \mathcal{J}_k^{B,\textrm{rej}}  \} $. }
		\STATE{\textbf{end while}}
		\STATE{\textbf{Until:} Achieve the stable matching $\vartheta^{*}_k$.}
		\STATE{\textbf{Sub-bands Assignment:} $\vartheta^{*}_k \rightarrow \boldsymbol{a}_k$.}

	\end{algorithmic}
	\label{Algorithm}
\end{algorithm}

\textbf{Preference of the device:} The preference function of device $j$ for sub-band $b$ can be defined as
\begin{equation}
    \theta_j(b) = \underbrace{\omega \log_2\left(1+ \frac{P_j^{k,b}g_j^{k,b}}{ \sum\limits_{k' \in \mathcal{K}, k' \neq k} \sum\limits_{j' \in \mathcal{J}, j' \neq j} P_{j'}^{k',b} g_{j'}^{k,b}+ \sigma^2}\right)}_{R_j^{k,b}}.
\end{equation}
The preference function of device $j \in \mathcal{J}_k$ in (34) indicates two facts: 1) the device's choice of sub-band only determines the transmission rate that can be achieved, which then determines the transmission delay and energy consumption when offloading a certain amount of data of its computation task to the associated UAV, as we can see in (11) and (12), and 2) the device would wish to offload a certain amount of its computation task to the associated UAV via the sub-band which can provide the highest transmission rate. 

\textbf{Preference of the UAV for its available sub-bands:} UAV $k$'s preference function for matching device $j\in \mathcal{J}_k$ with sub-band $b \in \mathcal{B}$ can be expressed as
\begin{equation}
    \theta_b(j) = \Phi_1  \underbrace{\omega \log_2\left(1+ \frac{P_j^{k,b}g_j^{k,b}}{ \sum\limits_{k' \in \mathcal{K}, k' \neq k} \sum\limits_{j' \in \mathcal{J}, j' \neq j} P_{j'}^{k',b} g_{j'}^{k,b}+ \sigma^2}\right)}_{R_j^{k,b}} - \underbrace{\sum\limits_{k' \in \mathcal{K}, k' \neq k} \Phi_j^{k',b} P_j^{k,b}g_j^{k',b} }_{\text{Cumulative interference to other UAVs}},
\end{equation}
where $\Phi_1$ and $\Phi_j^{k',b}$ are weighting parameters. The UAV will assign sub-band $b$ to device $j$ in order to maximize the achievable transmission rate and reduce cumulative interference to the other UAVs, as can be shown in (35). \\

\textit{Definition 2: A stable matching $\vartheta_k^{*}$ is achieved if there is no blocking pair $(j,b)$, where a pair $(j,b)$ is a blocking pair when $j \notin \vartheta_k(b), b \notin\vartheta_k(j)$, and $b\succ_{j}\vartheta_k(b)$ and $j\succ_{b}\vartheta_k(j)$.} \\

\begin{figure*}[t!]
\begin{mini!}[2]                 % maxi! = maximize
		{\boldsymbol{P}}                               % optimization variable
		{ \sum\limits_{ k \in \mathcal{K}} \sum\limits_{j \in \mathcal{J}} 	E_j^{j \rightarrow k, \textrm{trans}}(\boldsymbol{P}) }{\label{opt:P21}}{\textbf{P2.1:}} 
		\addConstraint{\frac{\beta_j^k}{ \omega \log_2\left(1+ \frac{P_j^{k,b} g_j^{k,b}}{ \sum\limits_{k' \in \mathcal{K}, k' \neq k} \sum\limits_{j' \in \mathcal{J}, j' \neq j} P_{j'}^{k',b} g_{j'}^{k,b}+ \sigma^2}\right)} \leq \varphi_j, \forall j \in \mathcal{J}_k, \forall k \in \mathcal{K},}
		\addConstraint{ 0 \leq P_j^{k,b} \leq P_j^{\mathbf{max}}, \forall j \in \mathcal{J}_k, \forall k \in \mathcal{K},}
\end{mini!}

	\hrulefill
\end{figure*}
The proposed game guarantees to converge to the stable matching since it is implemented identically to the
standard deferred acceptance algorithm \cite{roth2008deferred}. The pseudocode of the one-to-one matching game-based sub-band assignment algorithm is shown in Algorithm 1. Firstly, we acquire a set of devices $\mathcal{J}_k$, a set of sub-bands  $\mathcal{B}$, and initialize a set of unmatched devices $\mathcal{J}_k^{\textrm{un}}$, a set of prospective sub-bands for each device $\mathcal{B}_j$, a set of requested devices to each sub-band $\mathcal{J}_k^{b,\textrm{req}}$, and a set of rejected devices by each sub-band $\mathcal{J}_k^{b,\textrm{rej}}$. Every device builds its own preference list for all possible sub-bands (line 3) and then chooses the best sub-band $b$ (line 7) that can provide the highest transmission rate and sends the request to UAV $k$ in order to get access to that sub-band (line 8). When device $j$ selects sub-band $b$, the value of $q_{jb}$ is set to $1$, and if not, to $0$. After receiving requests from devices, the UAV updates the set of devices that have requested sub-band $b$ (line 11). Then, the UAV constructs the preference list of sub-band $b$ for all requested devices (lines 12-13). After that, the UAV will choose the best device for sub-band $b$ from the list of devices that have requested that sub-band, $\mathcal{J}_k^{b,\textrm{req}}$ (line 14), and assign the chosen device to sub-band $b$ (line 15). Then after, the set of rejected devices for sub-band $b$ is updated (line 16), and sub-band $b$ is deleted from the list of prospective sub-bands of its rejected devices (line 17). Finally, the set of unmatched devices is likewise updated based on the sets of rejected users for all sub-bands (line 19). The matching process is conducted iteratively until a stable match is established between both sides (i.e., devices and sub-bands). The process will stop when all devices are assigned to the sub-bands or there are no more sub-bands to send the access request to. Finally, the output of the one-to-one matching, $\vartheta^{*}_k$ is mapped to the sub-band assignment vector $\boldsymbol{a}_k$, i.e., $\vartheta^{*}_k \rightarrow \boldsymbol{a}_k$, (line 22).   \\

%\textcolor{blue}{
%\textbf{Theorem 1.} \textit{The stable matching $\vartheta^{*}_k$ can give local maximum for the sub-band assignment problem.}}

\textbf{Stage 2 (Power Control Problem)}. Utilizing the output of the proposed one-to-one matching game-based sub-band assignment algorithm that we presented in Algorithm 1, the power control problem can be expressed as \textbf{P2.1}. \\

\textbf{Theorem 1.} \textit{The objective function (36a) of the power control problem \textbf{P2.1}  is a concave function.}

\begin{IEEEproof}
Let us define
\begin{equation}
  L(\boldsymbol{P}_j^{k,b}) =  \frac{1}{ \omega \log_2\left(1+ \frac{g_j^{k,b}}{ \sum\limits_{k' \in \mathcal{K}, k' \neq k} \sum\limits_{j' \in \mathcal{J}, j' \neq j} P_{j'}^{k',b} g_{j'}^{k,b}+ \sigma^2}\right),}
\end{equation}
where $\boldsymbol{P}_j^{k,b} = \{P_{j'}^{1,b}, P_{j'}^{2,b}, \dots, P_{j'}^{K,b}\}$, $j' \in \mathcal{J}$. In accordance with the definition presented in (12), $E_j^{k,\textrm{trans}}$ which is the objective function (36a), is the perspective function of $ L(\boldsymbol{P}_j^{k,b})$, i.e., $E_j^{j\rightarrow k,\textrm{trans}}(\boldsymbol{P}) = \beta_j^kP_j^{k,b}L\left(\frac{\boldsymbol{P}_j^{k,b}}{P_j{k,b}}\right) $. Since the perspective function maintains concavity, if we can demonstrate that $L(\boldsymbol{P}_j^{k,b})$ is concave, then its perspective function $E_j^{j\rightarrow k,\textrm{trans}}(\boldsymbol{P})$  must also be concave. In order to keep things simple, we will demonstrate that $L(\boldsymbol{P}_{j'}^{k,b})$ is concave for a single variable. The case with multiple variables consists of a concave affine function and a single variable function, hence  if we are able to demonstrate that the perspective function is concave in the single variable scenario, then it will also be concave for multiple variables. Let us introduce $$M(P_{j'}^{k,b}) = \frac{1}{ \log_2\left(1+ \frac{1}{P_{j'}^{k',b}}\right)},  P_{j'}^{k',b}> 0$$ as the function of the single variable case of $ L(\boldsymbol{P}_j^{k,b})$. Then, the first-order derivative of $M(P_{j'}^{k,b})$ w.r.t $P_{j'}^{k',b},$ will be 
\begin{equation}
    \diff{M(P_{j'}^{k,b})}{P_{j'}^{k',b}} = \frac{1}{\ln2 P_{j'}^{k',b} ( P_{j'}^{k',b} +1) \log_2 \left(1 + \frac{1}{ P_{j'}^{k',b}}\right)^2.}
\end{equation}
From (38), we observe that $\diff{M(P_{j'}^{k,b})}{P_{j'}^{k',b}}>0$. Thus, $M(P_{j'}^{k,b})$ is a non-decreasing function of the transmit power profile $\boldsymbol{P}$. The second-order derivative is then
\begin{equation}
    \dfrac {\mathrm {d}^{2}M(P_{j'}^{k,b}) }{\mathrm {d}P_{j'}^{(k',b)2}} = \frac{2(P_{j'}^{k',b}+0.5)}{0.48\left((P_{j'}^{k',b})^2+P_{j'}^{k',b}\right)\log_2 \left(1 + \frac{1}{ P_{j'}^{k',b}}\right)^3} 
    \left[\frac{1}{P_{j'}^{k',b}+0.5} - \ln\left(1 + \frac{1}{P_{j'}^{k',b}}\right) \right].
\end{equation}
From (39), we can conclude that
\begin{equation}
    \frac{1}{P_{j'}^{k',b}+0.5} < \ln\left(1 + \frac{1}{P_{j'}^{k',b}}\right)
\end{equation}
when $P_{j'}^{k',b} > 0$. Therefore, $\dfrac {\mathrm {d}^{2}M(P_{j'}^{k,b}) }{\mathrm {d}P_{j'}^{(k',b)2}}< 0$, and $M(P_{j'}^{k,b})$ is concave. 
Additionally, constraint (36b) is concave, which can be shown following the same steps. 
\end{IEEEproof}

Thus, to make problem \textbf{P2.1} tractable, we first transform the problem into DC (i.e., difference of two convex functions) form. Following that, we develop a CCP (concave-convex procedure)-based technique to approach its stationary point, which is the optimal solution to the power control problem.

Firstly, the DC form of the constraint (36b) is 
\begin{equation}
     \Bigg[\underbrace{ \frac{\beta_j^k}{\omega \varphi_j}-\log_2\left(\sum\limits_{k \in \mathcal{K}} \sum\limits_{j \in \mathcal{J}}P_j^{k,b} g_j^{k,b}+ \sigma^2\right)}_{\hat{R}(\boldsymbol{P})} -
     \left(\underbrace{-\log_2\left(\sum\limits_{k' \in \mathcal{K}, k' \neq k} \sum\limits_{j' \in \mathcal{J}, j' \neq j} P_{j'}^{k',b} g_{j'}^{k,b}+ \sigma^2\right)}_{U(\boldsymbol{P})}\right)   \Bigg] \leq 0.
\end{equation}
Moreover, according to Theorem 1, the objective function (36a) is concave, thus, we can transform the objective function into a DC form, i.e., $0-\left(-E_j^{j\rightarrow k,\textrm{trans}}(\boldsymbol{P})\right)$. Finally, by approximating the concave parts of both the objective function and constraint in (41)  using the first-order Taylor approximation approach, we can convexify the objective function and the constraint. Thus, the following is the approximation function of $U(\boldsymbol{P})$ in (41)

\begin{equation}
    \bar{U}\left(\boldsymbol{P}^{(\hat{t}+1)}\right) =U\left(\boldsymbol{P}^{(\hat{t})}\right) + \nabla  U\left(\boldsymbol{P}^{(\hat{t})}\right)\left( \boldsymbol{P}^{(\hat{t}+1)} - \boldsymbol{P}^{(\hat{t})}\right) 
\end{equation}
where the subscript $\hat{t}$ is the current iteration and 

\begin{equation}
    \nabla  U\left(\boldsymbol{P}^{(\hat{t})}\right) = \frac{-\sum\limits_{k' \in \mathcal{K}, k' \neq k} \sum\limits_{j' \in \mathcal{J}, j' \neq j}g_{j'}^{k,b}}{\ln2\big(\log_2(\sum\limits_{k' \in \mathcal{K}, k' \neq k} \sum\limits_{j' \in \mathcal{J}, j' \neq j} P_{j'}^{k',b} g_{j'}^{k,b}+ \sigma^2)\big)}. 
\end{equation}
Consequently, the objective function's approximation function is defined as

\begin{equation}
\begin{split}
    \widehat{E}_j^{j\rightarrow k,\textrm{trans}}\left(\boldsymbol{P}^{(\hat{t}+1)}\right) &= - E_j^{j\rightarrow k,\textrm{trans}}\left(\boldsymbol{P}^{(\hat{t}+1)}\right) - \\
     & \  \ \ \nabla E_j^{j\rightarrow k,\textrm{trans}}\left(\boldsymbol{P}^{(\hat{t})}\right) \left( \boldsymbol{P}^{(\hat{t}+1)} - \boldsymbol{P}^{(\hat{t})}\right) 
\end{split}
\end{equation}
Finally, we can reformulate the power control problem as below
\begin{mini!}[2]                 % maxi! = maximize
		{\boldsymbol{P}}                               % optimization variable
		{ - \sum\limits_{ k \in \mathcal{K}} \sum\limits_{j \in \mathcal{J}} \widehat{E}_j^{j\rightarrow k,\textrm{trans}}(\boldsymbol{P}) }{\label{opt:P21}}{\textbf{P2.11:}}
		\addConstraint{\hat{R}(\boldsymbol{P}) - \bar{U}\left(\boldsymbol{P}\right) \leq 0, \forall j \in \mathcal{J}_k, \forall k \in \mathcal{K}, }
		\addConstraint{ 0 \leq P_j^{k,b} \leq P_j^{\mathbf{max}}, \forall j \in \mathcal{J}_k, \forall k \in \mathcal{K},}
\end{mini!}
where the objective function (45a) and constraint (45b) are convex, and the constraint (45c) is linear. Thus, problem \textbf{P2.11} is a convex problem. Therefore, we can solve it by using convex optimization techniques. The summary of the CCP-based power control algorithm is presented in Algorithm 2.

\begin{algorithm}[t]
	\caption{\strut CCP-Based Power Control Algorithm}
	\label{alg:profit}
	\begin{algorithmic}[1]
		\STATE{\textbf{Initialization:} Set $\hat{t}=0$, $\epsilon_1= 10^{-4}$, and find initial feasible solutions $(\boldsymbol{P}^{(0)})$;}
		
		\REPEAT
		
		\STATE{Solve the problem in (45) by using CVXPY toolkit and find the optimal transmit power profile $\boldsymbol{P}^{(\hat{t}+1)}$};
		\STATE{Update $\hat{t} =\hat{t} + 1$}; 
		\UNTIL{ $ \parallel \frac{ \widehat{E}_j^{j\rightarrow k,\textrm{trans}}(\boldsymbol{P}^{(\hat{t})}) - \widehat{E}_j^{j\rightarrow k,\textrm{trans}}(\boldsymbol{P}^{(\hat{t}+1)})}{ \widehat{E}_j^{j\rightarrow k,\textrm{trans}}(\boldsymbol{P}^{(\hat{t})}) } \parallel \ \leq \epsilon_1$};
		
		\STATE{Then, set $\boldsymbol{P}^{(\hat{t}+1)}$ as the desired solutions}.
	\end{algorithmic}
	\label{Algorithm}
\end{algorithm}

\subsection{UAV Deployment}

For a given $\{\boldsymbol{\beta, P, a, w, v, z}\}$, we can formulate the UAV deployment problem as
	\begin{mini!}[2]
	% maxi! = maximize
		{\boldsymbol{o}}                               % optimization variable
		{\mathbf{Q}(\boldsymbol{o}) }{\label{opt:P1}}{\textbf{P3:}} 
		\addConstraint{\text{(31c), (31m), (31n)}, }
	\end{mini!}
where $\mathbf{Q}(\boldsymbol{o}) = \sum\limits_{j \in \mathcal{J}_k}\sum\limits_{k \in \mathcal{K}}E_j^{j \rightarrow k, \textrm{trans}}+ \sum\limits_{k \in \mathcal{K}}\bigg(\sum\limits_{k' \in \mathcal{K}, k' \neq k} E^{k \rightarrow k',\textrm{trans}}+\sum\limits_{s \in \mathcal{S}}  E^{k \rightarrow s, \textrm{trans}}\bigg)$. However, problem \textbf{P3} is non-convex due to the non-convex objective function and constraint (31c). Thus, we use a successive convex approximation (SCA) approach to address the formulated UAV deployment problem and to achieve a locally optimal solution. SCA iteratively approximates every non-convex function in the optimization problem with a convex function. Then, the approximated convex problem is solved via standard optimization techniques. 

We first introduce sets of auxiliary variables $\boldsymbol{\lambda}_{j,k} = \{ \lambda_{j,k}, \forall j \in \mathcal{J}_k, \forall k \in \mathcal{K} \}$, $\boldsymbol{\lambda}_{k,k'} =  \{ \lambda_{k,k'}, \forall k,k' \in \mathcal{K} \}$, and  $\boldsymbol{\lambda}_{k,s} = \{ \lambda_{k,s}, \forall k \in \mathcal{K}, \forall s \in \mathcal{S} \}$ in order to replace non-linear inequality constraint (31c) with four inequality constraints as follows

\begin{equation}
 \frac{\beta_j^k}{R_j^k} \leq \lambda_{j,k}, \forall j \in \mathcal{J}_k, \forall k \in \mathcal{K},
\end{equation} 

\begin{equation}
\frac{\beta^{k\rightarrow k'}}{R^{k \rightarrow k'}} \leq \lambda_{k,k'},  \forall k,k' \in \mathcal{K},
\end{equation}

\begin{equation}
\frac{\beta^{k\rightarrow s}}{R^{k \rightarrow s}} \leq \lambda_{k,s},  \forall k \in \mathcal{K}, \forall s \in \mathcal{S},
\end{equation}

\begin{equation} 
 \lambda_{j,k} + \lambda_{k,k'} +   \lambda_{k,s} \leq  \varphi_j, \forall j \in \mathcal{J}_k, \forall k,k' \in \mathcal{K}, \forall s \in \mathcal{S}.
\end{equation}
Then, we can reformulate problem \textbf{P3} as shown in \textbf{P3.1}.
\begin{figure*}[t!]
\begin{mini!}[2]                 % maxi! = maximize
		{\boldsymbol{o}}                               % optimization variable
		{\sum\limits_{j \in \mathcal{J}_k}\sum\limits_{k \in \mathcal{K}}E_j^{j \rightarrow k, \textrm{trans}}+ \sum\limits_{k \in \mathcal{K}}\bigg(\sum\limits_{k' \in \mathcal{K}, k' \neq k} E^{k \rightarrow k',\textrm{trans}}+ \sum\limits_{s \in \mathcal{S}}  E^{k \rightarrow s, \textrm{trans}}\bigg) }{\label{opt:P1}}{\textbf{P3.1:}} 
		\addConstraint{\underbrace{ \frac{\beta_j^k}{\omega \lambda_{j,k}}-\log_2\left(\sum\limits_{k \in \mathcal{K}} \sum\limits_{j\in \mathcal{J}} \frac{P_{j}^{k,b} g_0}{(d_j^k)^{2} e^{i_b(f)d_j^k}} + \sigma^2\right)}_{ \emph{N}_1(\boldsymbol{o})} - }\nonumber
		\addConstraint{ \left(\underbrace{-\log_2\left(\sum\limits_{k' \in \mathcal{K}, k' \neq k} \sum\limits_{j' \in \mathcal{J}, j' \neq j} \frac{P_{j'}^{k',b}g_0}{(d_{j'}^k)^{2} e^{i_b(f)d_{j'}^k}}+ \sigma^2\right)}_{\emph{N}_2(\boldsymbol{o})}\right)  \leq 0, \forall j \in \mathcal{J}_k, \forall k \in \mathcal{K},}
		\addConstraint{ \text{(48), and (49)}, }
		\addConstraint{ \text{(31m), (31n), (50)}, }
\end{mini!}
	\hrulefill
\end{figure*}
Due to the logarithmic terms in $R_j^k$ (i.e., $\emph{N}_1(\boldsymbol{o})$ and $\emph{N}_2(\boldsymbol{o})$), $R^{k \rightarrow k'}$, and $R^{k \rightarrow s}$, problem \textbf{P3.1} is non-convex. Thus, as shown in (52), we first convexify $\emph{N}_1(\boldsymbol{o})$ by introducing its convex lower bound function, $\hat{\emph{N}}_1(\boldsymbol{o})$, based on the first-order Taylor approximation at the given location of UAV $k$ at $t$-th iteration, $ \boldsymbol{o}_k(t)$. In the same way, $\emph{N}_2(\boldsymbol{o})$), $R^{k \rightarrow k'}$, and $R^{k \rightarrow s}$ can be convexified as shown in (53), (55) and (57), respectively.

\begin{figure*}
\begin{equation}
\begin{split}
    \hat{\emph{N}}_1(\boldsymbol{o}) &= \left( \frac{\beta_j^k}{\omega \lambda_{j,k}}-\log_2\bigg(\sum\limits_{k \in \mathcal{K}} \sum\limits_{j\in \mathcal{J}} \frac{P_{j}^{k,b} g_0}{(h_k^2+ \parallel \boldsymbol{o}_k(t) - \boldsymbol{\ddot{c}}_j \parallel^2 ) e^{i_b(f)(h_k^2+\parallel \boldsymbol{o}_k(t) - \boldsymbol{\ddot{c}}_j\parallel^2 )^{1/2}}} + \sigma^2\bigg) \right) + \bigg( \parallel \boldsymbol{o}_k - \boldsymbol{\ddot{c}}_j\parallel^2   \\
   & \  -\parallel \boldsymbol{o}_k (t) - \boldsymbol{\ddot{c}}_j\parallel^2     \bigg) \frac{\sum\limits_{k \in \mathcal{K}} \sum\limits_{j\in \mathcal{J}}\frac{P_{j}^{k,b} g_0 \left[ \frac{1}{2}(h_k^2+\parallel \boldsymbol{o}_k(t) - \boldsymbol{\ddot{c}}_j\parallel^2 )^{1/2}e^{i_b(f)(h_k^2+\parallel \boldsymbol{o}_k(t) - \boldsymbol{\ddot{c}}_j\parallel^2)^{1/2} } i_b(f) + e^{i_b(f)(h_k^2+\parallel \boldsymbol{o}_k(t) - \boldsymbol{\ddot{c}}_j\parallel^2 )^{1/2}}     \right]}{\left((h_k^2+\parallel \boldsymbol{o}_k(t) - \boldsymbol{\ddot{c}}_j\parallel^2 ) e^{i_b(f)(h_k^2+\parallel \boldsymbol{o}_k(t) - \boldsymbol{\ddot{c}}_j\parallel^2)^{1/2} } \right)^2}}{\ln2 \bigg(\sum\limits_{k \in \mathcal{K}} \sum\limits_{j\in \mathcal{J}} \frac{P_{j}^{k,b} g_0}{(h_k^2+\parallel \boldsymbol{o}_k(t) - \boldsymbol{\ddot{c}}_j\parallel^2 ) e^{i_b(f)(h_k^2+\parallel \boldsymbol{o}_k(t) - \boldsymbol{\ddot{c}}_j\parallel^2 )^{1/2}}} + \sigma^2\bigg) } 
\end{split} 
\end{equation}

\begin{equation}
 \begin{split}
  &\hat{\emph{N}}_2(\boldsymbol{o}) =   -\log_2\left(\sum\limits_{k' \in \mathcal{K}, k' \neq k} \sum\limits_{j' \in \mathcal{J}, j' \neq j} \frac{P_{j'}^{k',b}g_0}{(h_k^2 + \ddot{n}_k) e^{i_b(f)(h_k^2 + \ddot{n}_k)^{1/2}}}+ \sigma^2\right), 
\end{split}  
\end{equation}
where
\begin{equation}
 \begin{split}
 & \ddot{n}_k \leq \parallel \boldsymbol{o}_k (t) - \boldsymbol{\ddot{c}}_{j'} \parallel ^2  + 2 ( \boldsymbol{o}_k (t) - \boldsymbol{\ddot{c}}_{j'})^T (\boldsymbol{o}_k - \boldsymbol{o}_k (t) ), \forall k' \in \mathcal{K}, k'\neq k, \forall j' \in \mathcal{J}_k, j'\neq j, 
\end{split}  
\end{equation}

\begin{equation}
   \begin{split}
       \ddot{R}^{k \rightarrow k'} =   B_{\textrm{mm}}^{k\rightarrow k'} \log_2\left(1 +  \frac{P^{k \rightarrow k'} g_k^{\textrm{tx}} g_{k'}^{\textrm{rx}}L_r}{t_nHB_{\textrm{mm}}^{k \rightarrow k'}}   \left( \frac{c^2}{16\pi^2 \bigg( (h_k - h_{k'})^2 + \hat{n}_k   \bigg) (f_c^{\textrm{mm}})^2}\right)\right), \forall k, k' \in \mathcal{K},
   \end{split}
\end{equation}
where 

\begin{equation}
  \begin{split}
       \hat{n}_k \geq  \parallel  \boldsymbol{o}_k(t) - \boldsymbol{o}_{k'}(t) \parallel^2 + 2(  \boldsymbol{o}_k(t) - \boldsymbol{o}_{k'}(t))^T( \boldsymbol{o}_k - \boldsymbol{o}_{k'})     
   \end{split}
\end{equation}

\begin{equation}
    \begin{split}
       \ddot{R}^{k \rightarrow s} =  B_{\textrm{mm}}^{k\rightarrow s} \log_2\left(1 +  \frac{P^{k \rightarrow s} g_k^{\textrm{tx}} g_{s}^{\textrm{rx}}L_r}{t_nHB_{\textrm{mm}}^{k \rightarrow s}}   \left( \frac{c^2}{16\pi^2 \bigg( (h_k - h_{s})^2 + \Tilde{n}_k   \bigg) (f_c^{\textrm{mm}})^2}\right)\right), \forall k \in  \mathcal{K}, \forall s \in  \mathcal{S},
   \end{split}
\end{equation}
where
\begin{equation}
 \begin{split}
  & \Tilde{n}_k \geq \parallel \boldsymbol{o}_k (t) - \boldsymbol{o}_{s} \parallel ^2  + 2 ( \boldsymbol{o}_k (t) - \boldsymbol{o}_{s})^T (\boldsymbol{o}_k - \boldsymbol{o}_k (t) ),\forall k \in \mathcal{K}, \forall s \in \mathcal{S},  
\end{split} 
\end{equation}
 \hrulefill
\end{figure*}

 Finally, we can approximate the non-convex problem \textbf{P3.1} as a convex problem as the following
	\begin{mini!}[2]
	% maxi! = maximize
		{\boldsymbol{o,\lambda, \ddot{n},  \hat{n},  \Tilde{n}}}                               % optimization variable
		{\hat{\mathbf{Q}}(\boldsymbol{o, \lambda,\ddot{n},  \hat{n},  \Tilde{n}}) }{\label{opt:P1}}{\textbf{P3.2:}} 
		\addConstraint{\hat{\emph{N}}_1(\boldsymbol{o}) - \hat{\emph{N}}_2(\boldsymbol{o}) \leq 0, \forall j \in \mathcal{J}_k, \forall k \in \mathcal{K}, }
		\addConstraint{\frac{\beta^{k\rightarrow k'}}{\ddot{R}^{k \rightarrow k'}} \leq \lambda_{k,k'},  \forall k,k' \in \mathcal{K},}
		\addConstraint{\frac{\beta^{k\rightarrow s}}{\ddot{R}^{k \rightarrow s}} \leq \lambda_{k,s},  \forall k \in \mathcal{K}, s \in \mathcal{S},}
		\addConstraint{ \text{(30m), (30n), (50), (54), (56), (58)}, }
	\end{mini!}
where $\hat{\mathbf{Q}}(\boldsymbol{o, \lambda,  \ddot{n},  \hat{n},  \Tilde{n} }) = \sum\limits_{j \in \mathcal{J}_k}\sum\limits_{k \in \mathcal{K}} \frac{P_j^{k,b} \beta_j^k}{\ddot{R}_j^k}+ \sum\limits_{k \in \mathcal{K}}\bigg(\sum\limits_{k' \in \mathcal{K}, k' \neq k} \frac{P^{k \rightarrow k'}\beta^{k\rightarrow k'}}{ \ddot{R}^{k \rightarrow k'}}+ \sum\limits_{s \in \mathcal{S}}\frac{P^{k \rightarrow s}\beta^{k\rightarrow s}}{ \ddot{R}^{k \rightarrow s}} \bigg)$. As problem \textbf{P3.2} is a convex problem, we can solve it using convex optimization techniques. The summary of the SCA-based optimal UAVs deployment algorithm is presented in Algorithm 3.  

\begin{algorithm}[h]
	\caption{\strut UAV Deployment Algorithm}
	\label{alg:profit}
	\begin{algorithmic}[1]
		\STATE{\textbf{Initialization:} Set $t=0$, $\epsilon_2= 10^{-4}$, and find initial feasible solutions $(\boldsymbol{o}^{(0)}, \boldsymbol{\lambda}^{(0)}, \boldsymbol{\ddot{n}}^{(0)}, \boldsymbol{\hat{n}}^{(0)}, \boldsymbol{\Tilde{n}}^{(0)})$;}
		
		\REPEAT
		
		\STATE{Solve the problem in (59) by using CVXPY toolkit and find the optimal location of UAVs and auxiliary variables, $\boldsymbol{o}^{(t+1)}$, $\boldsymbol{\lambda}^{(t+1)}$, $\boldsymbol{\ddot{n}}^{(t+1)}$,  $\boldsymbol{\hat{n}}^{(t+1)}$, and  $\boldsymbol{\Tilde{n}}^{(t+1)}$ };
		\STATE{Update $t = t + 1$}; 
		\UNTIL{ $ \parallel \frac{ \hat{\mathbf{Q}}^{(t)} - \hat{\mathbf{Q}}^{(t+1)}}{ \hat{\mathbf{Q}}^{(t)}} \parallel \ \leq \epsilon_2$};
		
		\STATE{Then, set $\boldsymbol{o}^{(t+1)}$, $\boldsymbol{\lambda}^{(t+1)}$, $\boldsymbol{\ddot{n}}^{(t+1)}$,  $\boldsymbol{\hat{n}}^{(t+1)}$, and  $\boldsymbol{\Tilde{n}}^{(t+1)}$ as the desired solution}.
	\end{algorithmic}
	\label{Algorithm}
\end{algorithm}

\subsection{UAV Task Offloading Decision}
For a given $\{\boldsymbol{\beta, o, P, a}\}$, we can formulate the UAV task offloading decision problem as
	\begin{mini!}[2]                 % maxi! = maximize
		{\boldsymbol{w, v, z}}                               % optimization variable
		{\mathbf{Q}(\boldsymbol{w, v, z})}{\label{opt:P1}}{\textbf{P4:}} 
		\addConstraint{\text{(31c), (31h), (31j)-(31l)}, }
  \vspace{-0.1in}
	\end{mini!}
Problem \textbf{P4} is non-convex and is a combinatorial problem. As a result, we propose to use the BSUM method to address problem \textbf{P4} \cite{hong2015unified}. BSUM is a method for addressing non-convex and non-smooth optimization problems by splitting the problem into manageable subproblems. Using the BSUM approach, the decision variables $\boldsymbol{w, v, z}$ are updated consecutively in order to minimize the upper bound of the objective function. Additionally, BSUM can ensure convergence to the stationary points of the objective function in (60a). To use the BSUM technique, we first relax the binary constraints (31j)-(31l) and replace them with continuous ones. Then, we can introduce the feasible sets of 
$\boldsymbol{w}$, $\boldsymbol{v}$, and $\boldsymbol{z}$ as the following 
\begin{align*}    
\mathcal{W} \triangleq & \{ \boldsymbol{w}:l_j^{k, \textrm{Remote}}  \leq \varphi_j, w_j^{k} + \sum_{\substack{k'\in \mathcal{K},  k'\neq k}} v_j^{k \rightarrow k'} + \sum_{s \in \mathcal{S}} z_j^{k\rightarrow s} = 1, \\
 & w_j^{k} \in [0,1], \forall j \in \mathcal{J}_k, \forall k \in \mathcal{K}\},
\end{align*}

\begin{align*}    
\mathcal{V} \triangleq & \{\boldsymbol{v}: l_j^{k, \textrm{Remote}}  \leq \varphi_j, w_j^{k} + \sum_{\substack{k'\in \mathcal{K},  k'\neq k}} v_j^{k \rightarrow k'} + \sum_{s \in \mathcal{S}} z_j^{k\rightarrow s} = 1, \\
& v_j^{k \rightarrow k'} \in [0,1], \forall j \in \mathcal{J}_k, \forall k,k' \in \mathcal{K}\},
\end{align*}
\begin{align*}    
\mathcal{Z} \triangleq & \{\boldsymbol{z}: l_j^{k, \textrm{Remote}}  \leq \varphi_j, w_j^{k} + \sum_{\substack{k'\in \mathcal{K}, k'\neq k}} v_j^{k \rightarrow k'} + \sum_{s \in \mathcal{S}} z_j^{k\rightarrow s} = 1, \\
& z_j^{k \rightarrow s} \in [0,1], \forall j \in \mathcal{J}_k, \forall k \in \mathcal{K}, \forall s \in \mathcal{S}\},
\end{align*}

\begin{algorithm}[t!]
	\caption{\strut BSUM-Based UAV Task Offloading Decision Algorithm}
	\label{alg:profit}
	\begin{algorithmic}[1]
		\STATE{\textbf{Initialization:} Set $\ddot{t}=0$, $\epsilon_3= 10^{-4}$, and initial solutions $(\boldsymbol{w}^{(0)}, \boldsymbol{v}^{(0)}, \boldsymbol{z}^{(0)})$;}
		
		\REPEAT
		
		\STATE{Choose index set $\mathcal{M}$};
		\STATE{Let $\boldsymbol{w}^{(\ddot{t}+1)}_m \in \underset{\boldsymbol{w}_m }{\mathrm{argmin}} \ 
			\mathbf{Q}_m\big(\boldsymbol{w}_m; \boldsymbol{w}^{(\ddot{t})},  \boldsymbol{v}^{(\ddot{t})}, \boldsymbol{z}^{(t)}\big)$};
		\STATE{Set $\boldsymbol{w}_n^{(\ddot{t}+1)} = \ \boldsymbol{w}_n^{\ddot{t}}$, $\forall n \notin \mathcal{M}$};
		\STATE{Find  $\boldsymbol{v}_m^{(\ddot{t}+1)}$, and $\boldsymbol{z}_m^{(\ddot{t}+1)}$ by addressing (63) and (64)};
		\STATE{Update $\ddot{t} = \ddot{t} + 1$}; 
		\UNTIL{ $ \parallel \frac{\mathbf{Q}_m^{(\ddot{t})} -   \ \mathbf{Q}_m^{(\ddot{t}+1)}}{\mathbf{Q}_m^{(\ddot{t})}} \parallel \ \leq \epsilon_3$}
		
		\STATE{Then, set $\big(\boldsymbol{w}_m^{(\ddot{t}+1)}$, $\boldsymbol{v}_m^{(\ddot{t}+1)}$, $\boldsymbol{z}_m^{(\ddot{t}+1)} \big)$ as the desired solution}.
	\end{algorithmic}
	\label{Algorithm}
\end{algorithm}

Finally, we establish the proximal upper bound function of the objective function (60a) for each iteration  $\ddot{t}$, $\forall m \in \mathcal{M}$, where $\mathcal{M}$ is the index set, as shown below %optimization problem as
%	\begin{mini!}[2]                 % maxi! = %maximize
%		{\boldsymbol{w\in \mathcal{W}, v \in \mathcal{V}, z\in \mathcal{Z}}}                      %         % optimization variable
%		{\mathbf{Q}(\boldsymbol{w, v, z})}{\label{opt:P1}}{\textbf{P4.1:}}
%	\end{mini!}
%Following that, we establish the proximal upper-bound function of the objective function (61a) for each iteration $\ddot{t}$, $\forall m \in \mathcal{M}$, where $\mathcal{M}$ is the index set, as shown below

\begin{equation} 
\begin{split}
\mathbf{Q}_m(\boldsymbol{w}_m; \boldsymbol{w}^{\ddot{t}}, \boldsymbol{v}^{\ddot{t}}  \boldsymbol{z}^{\ddot{t}}) =  \mathbf{Q}(\boldsymbol{w}_m; \widehat{\boldsymbol{w}},\widehat{\boldsymbol{v}}, \widehat{\boldsymbol{z}}) + \frac{\mu_m}{2} \parallel ( \boldsymbol{w}_m
- \widehat{\boldsymbol{w}}) \parallel^2
\end{split}
\end{equation} 
where the quadratic penalty term helps to convexify the proximal upper-bound function, and $\mu_m$ is a positive penalty parameter that can be used for the other vectors of the variables $\boldsymbol{v}$, and $\boldsymbol{z}$, respectively. Additionally, the proximal upper-bound function (61) contains distinct minimizer vectors $\widehat{\boldsymbol{w}},\widehat{\boldsymbol{v}}$, and $\widehat{\boldsymbol{z}}$ with respect to $\boldsymbol{w}$, $\boldsymbol{v}$, and $\boldsymbol{z}$  at each iteration $\ddot{t}$, which are taken into account as the solution of the preceding iteration $(\ddot{t}-1)$. The solution at  iteration $(\ddot{t} + 1)$ is then obtained by solving the subproblems 

\begin{align}
\boldsymbol{w}^{(\ddot{t}+1)}_m  & \in \underset{\boldsymbol{w}_m }{\mathrm{argmin}} \ \
\mathbf{Q}_m\bigg(\boldsymbol{w}_m; \boldsymbol{w}^{(\ddot{t})},  \boldsymbol{v}^{(\ddot{t})}, \boldsymbol{z}^{(\ddot{t})} \bigg),  \\
\boldsymbol{v}^{(\ddot{t}+1)}_m  & \in \underset{\boldsymbol{v}_m }{\mathrm{argmin}} \ \
\mathbf{Q}_m\bigg(\boldsymbol{v}_m; \boldsymbol{v}^{(\ddot{t})},  \boldsymbol{w}^{(\ddot{t}+1)}, \boldsymbol{z}^{(\ddot{t})} \bigg),  \\
\boldsymbol{z}_m^{(\ddot{t}+1)}  & \in \underset{\boldsymbol{z}_m }{\mathrm{argmin}}  \ \
\mathbf{Q}_m\bigg(\boldsymbol{z}_m; \boldsymbol{z}^{(\ddot{t})}, \boldsymbol{w}^{(\ddot{t}+1)}, \boldsymbol{v}^{(\ddot{t}+1)}\bigg). 
\end{align} 
Subproblems (62), (63), and (64) can  be solved by using convex optimization techniques. A summary of our proposed  BSUM-based UAVs tasks offloading decision algorithm is presented in Algorithm 4. 
\begin{algorithm}[t!]
	\caption{\strut Joint Task Offloading, Sub-band Assignment, Power Control, and UAV Deployment Algorithm}
	\label{alg:profit}
	\begin{algorithmic}[1]
		\STATE{\textbf{Initialization:} Set $\Tilde{t}=0$, $\epsilon_4= 10^{-4}$, and initial solutions $(\boldsymbol{\beta}^{(0)}, \boldsymbol{a}^{(0)}, \boldsymbol{P}^{(0)}, \boldsymbol{o}^{(0)}, \boldsymbol{w}^{(0)} \boldsymbol{v}^{(0)}, \boldsymbol{z}^{(0)})$;}
		
		\REPEAT
		
		\STATE{Solve device task offloading problem \textbf{P1} at the given $(\boldsymbol{a}^{(\Tilde{t})}, \boldsymbol{P}^{(\Tilde{t})}, \boldsymbol{o}^{(\Tilde{t})}, \boldsymbol{w}^{(\Tilde{t})} \boldsymbol{v}^{(\Tilde{t})}, \boldsymbol{z}^{(\Tilde{t})})$ by using CVXPY toolkit};
		\STATE{Solve sub-band assignment and transmit power control problem at the given $(\boldsymbol{\beta}^{(\Tilde{t}+1)}, \boldsymbol{o}^{(\Tilde{t})}, \boldsymbol{w}^{(\Tilde{t})} \boldsymbol{v}^{(\Tilde{t})}, \boldsymbol{z}^{(\Tilde{t})})$ by using Algorithm 1 and Algorithm 2};
		\STATE{Solve UAV deployment problem at the given $(\boldsymbol{\beta}^{(\Tilde{t}+1)}, \boldsymbol{a}^{(\Tilde{t}+1)}, \boldsymbol{P}^{(\Tilde{t}+1)}, \boldsymbol{w}^{(\Tilde{t})} \boldsymbol{v}^{(\Tilde{t})}, \boldsymbol{z}^{(\Tilde{t})})$ using Algorithm 3 };
		\STATE{Solve UAV task offloading decision problem at the given $(\boldsymbol{\beta}^{(\Tilde{t}+1)}, \boldsymbol{a}^{(\Tilde{t}+1)}, \boldsymbol{P}^{(\Tilde{t}+1)}, \boldsymbol{o}^{(\Tilde{t}+1)} )$ by using Algorithm 4};
		\STATE{Update $\Tilde{t} = \Tilde{t} + 1$}; 
		\UNTIL{ $ \parallel \frac{\mathbf{Q}^{(\Tilde{t})} -   \ \mathbf{Q}^{(\Tilde{t}+1)}}{\mathbf{Q}^{(\Tilde{t})}} \parallel \ \leq \epsilon_4$}
		
		\STATE{Set $ \big(\boldsymbol{\beta}^{(\Tilde{t}+1)}, \boldsymbol{a}^{(\Tilde{t}+1)}, \boldsymbol{P}^{(\Tilde{t}+1)}, \boldsymbol{o}^{(\Tilde{t}+1)}, \boldsymbol{w}^{(\Tilde{t}+1)}, \boldsymbol{v}^{(\Tilde{t}+1)}, \boldsymbol{z}^{(\Tilde{t}+1)}  \big)$ as the desired solution}.
	\end{algorithmic}
	\label{Algorithm}
\end{algorithm}

\subsection{Complexity of Joint Task Offloading, Sub-band Assignment, Power Control, and UAV Deployment Algorithm}

Our proposed joint task offloading, sub-band assignment, power control, and UAV deployment algorithm is summarized in Algorithm 5. The algorithm follows an alternating optimization paradigm that calls for resolving subproblems in (32), (33), (46), and (60) repeatedly prior to convergence. At each iteration, the complexity of the device task offloading decision is $\mathcal{O}(J_k K )$. Then, the complexity of achieving the stable matching in a one-to-one matching game-based sub-channel assignment algorithm is $\mathcal{O}(J_kB)$. The computational complexity of the proposed CCP-based power control algorithm is $\mathcal{O}((J_k K )^3 (2J_k K))$ \cite{bandi2019joint}. The complexity of the SCA-based UAV deployment algorithm in Algorithm 3 is $\mathcal{O}((K)^{3.5})$\cite{li2019energy}. The complexity of the BSUM-based algorithm to solve the UAV task offloading decision problem in (60) is $\mathcal{O}((KS+K^2)^{3.5})$. Therefore, at each iteration, the complexity of the proposed joint task offloading, sub-band assignment, power control, and UAV deployment algorithm presented in Algorithm 5 is $\mathcal{O}(J_k K+J_kB+ (J_k K )^3 (2J_k K)+(KS+K^2)^{3.5})$.             

\section{Simulation Results}
\label{simu}

\subsection{Evaluation Methodology}

To evaluate the proposed solution, we consider wireless devices distributed within an area of $600$ m × $600$ m. To provide computing services to the devices, $4$ MEC-enabled UAVs hover at an altitude of $50$ m. Additionally, $2$ LEO satellites at an altitude of $[780, 800]$ km are taken into consideration to execute the devices' tasks that the UAVs cannot handle; their locations are assumed to be unchanged during the simulation. The data size of the task, $A_j$, is selected from a uniform distribution on $[0.1, 0.5]$ Mbits. Furthermore, the required CPU cycles to compute a bit of data, $\alpha_j$, is also selected from a uniform distribution on $[10, 50]$ Cycles. The rest of the simulation parameters are shown in Table I. We use Python programming language to conduct simulation, and all of the proposed algorithms are executed on the PC with Intel(R) Core(TM) i5-8500 CPU @3.00GHz 3.00 GHz, 32.0 GB RAM, and NVIDIA GeForce GTX 1660 Ti. As a basis for comparison, we use two baseline schemes proposed in the recent literature  \cite{zhu2020joint} and \cite{mao2020joint}, namely: 1) \emph{All local computing} scheme where devices compute their tasks locally, and 2) \emph{No UAVs collaboration} scheme in which the computation capacity of the UAV is not sufficient to execute the offloaded tasks of its associated devices, the UAV directly transferred its devices' tasks to LEO satellites using mmWave backhaul links without checking its neighboring UAVs which have sufficient computation to execute its computation tasks. The results shown in this work are the averages of 100 simulations.   
\begin{table}[t!]
	%	\caption{Summary of Key Notations}	
	\caption{Simulation Parameters.}
	\textbf{\label{tab:table_simulation}} 
	\renewcommand\arraystretch{1}
	\begin{center}
		\begin{tabular}{|m{1.5cm}|m{2cm}||m{1.5cm}|m{2cm}|}
			%		\begin{tabular}{|p{1.5cm}|p{6.5cm}||p{2cm}|p{6.5cm}|}
			\hline
			\hfil \textbf{Parameter} & \hfil \textbf{Value} & \hfil \textbf{Parameter} & \hfil \textbf{Value} \\ \hline \hline
			\ \hfil $B$ & 25 &  \hfil\hfil $\varphi_j$ & 500 ms \\ \hline
			\ \hfil $g_0$ & -20 dB  & \hfil $P_j^{\mathbf{max}}$ &  23 dBm \\ \hline
			\ \hfil $\sigma^2$ & -174 dBm  & \hfil\hfil $f_j$ & 0.01 MHz  \\ \hline
			\ \hfil $\kappa_j, \kappa$ & 1 $\times$ $10^{-10}$  & \hfil $\omega$ &  5 $\times$  $10^{2}$ Hz  \\ \hline
			\ \hfil $i_b(f)$ & 0.005 \textcolor{blue}{\cite{xu2021joint}}& \hfil $B^{k \rightarrow k'}$ & 1.7 MHz  \\ \hline
			\ \hfil  $P^{k \rightarrow k'}$ & 30 dBm & \hfil  \hfil $g_k^{\textrm{tx}}, g_{k'}^{\textrm{rx}}$ & 41 dB \\ \hline
			\ \hfil $L_r$ & -23 dB & \hfil $H$  & 300 K \\ \hline 
			\ \hfil $F_k^{\mathbf{max}}$ & 3.5 MHz    & \hfil $P^{k \rightarrow k'}$ & 30 dBm \\ \hline 
			\ \hfil $B^{k \rightarrow k'}$ & 1.7 MHz & \hfil $B_{\textrm{mm}}^{k \rightarrow s}$ & 1.8 MHz \\ \hline
			
			\ \hfil $f_c^{\textrm{mm}}$ & 28 GHz & \hfil $P^{k \rightarrow s}$ & 30 dBm \\ \hline 
			%\ \hfil $\varpi_k$ &  5 $\times$  10^{27} & \hfil $\varpi_j$ &  5 $\times$ 10^{27}  \\ \hline 
		\end{tabular}
         
		\label{tab1}
	\end{center}
\end{table}

    \begin{figure}[t]
    \centering
    \captionsetup{justification = centering}
    \includegraphics[width=0.5\linewidth]{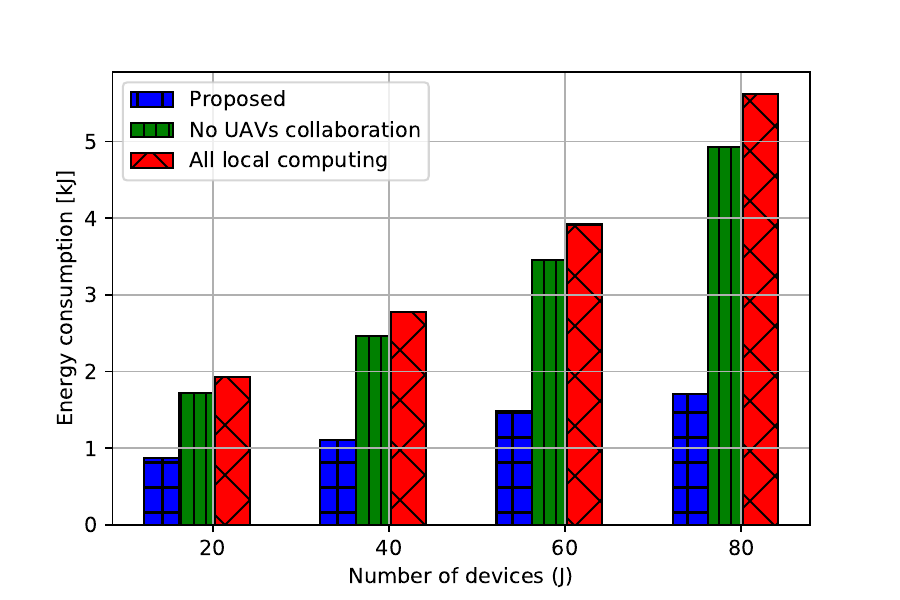}
    \caption{Energy consumption vs. number of devices for proposed, local computing only and without UAVs collaboration.}
    \label{enernocollab}
    \end{figure}
    
\subsection{Energy Consumption Analysis}

Fig.~\ref{enernocollab} shows the energy consumption as a function of the number of devices in the system, obtained using the proposed algorithm with two state-of-the-art schemes in the literature. The figure demonstrates that compared to other schemes, the total energy consumption at UAVs and devices to accomplish the execution of devices' computation tasks under our proposed scheme is the lowest in every network size. The figure also shows that as the network size increases, the performance gap between the proposed algorithm and two state-of-the-art schemes widens. As a result, we conclude that the proposed algorithm is also appropriate for large-scale networks. Finally, we see how crucial collaboration among UAVs is to the integrated SAG networks by analyzing the energy consumption under the \emph{No UAVs collaboration} scheme \cite{mao2020joint,  hu2023joint} in the figure. In contrast to the proposed algorithm, the \emph{No UAVs collaboration} scheme results in higher energy consumption since satellites are farther away from the UAV than its neighboring UAVs, which results in higher transmission energy (i.e., UAV-to-satellite transmission energy) than the UAV-to-UAV transmission energy.

  \begin{figure}[h!]
  \centering
 \captionsetup{justification = centering}
 \includegraphics[width=0.5\linewidth]{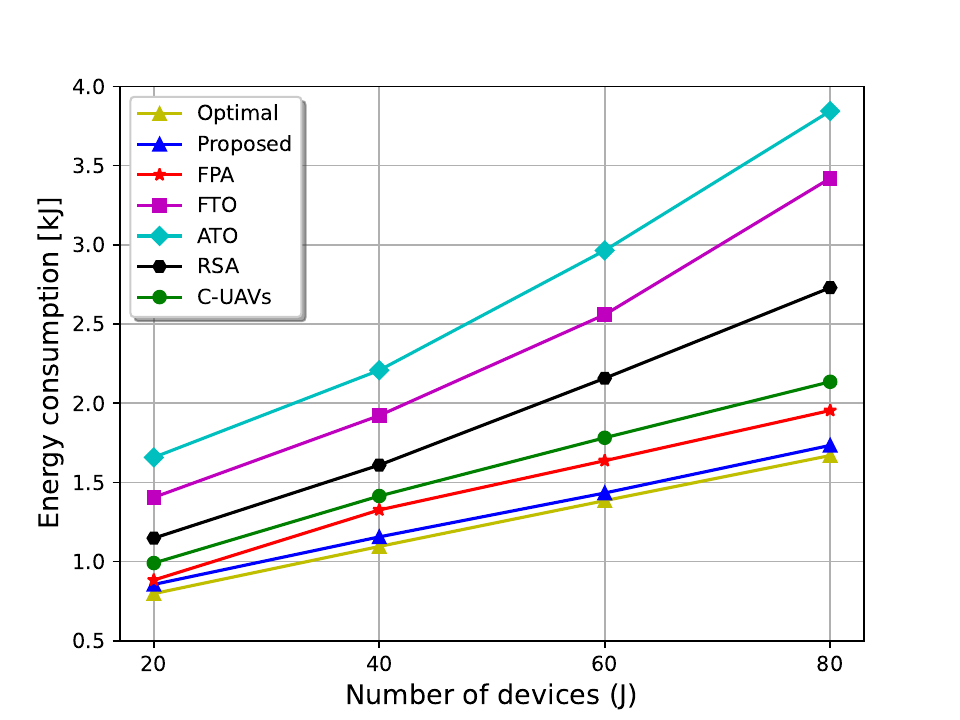}
\caption{Energy consumption as a function of the number of devices for variants of the proposed scheme.}
\label{devi}
\end{figure}

In order to evaluate the importance of different decision variables in minimizing the energy consumption, in what follows we consider the following variants of the proposed solution:

\begin{itemize}
    \item \emph{Centered UAVs (C-UAVs):} Each UAV is deployed at the center of its associated devices, i.e., at the center of each cluster, which we established via the K-means algorithm. At the same time, sub-band assignment, power control, and UAV task offloading problems are solved via the proposed Algorithm 1, Algorithm 2, and Algorithm 4.
    
    \item \emph{All tasks offloading (ATO)}: In this variant, devices offload all of their computation tasks to their associated UAVs to perform remote computing. Sub-band assignment, power control, UAV deployment, and UAV task offloading problems are solved via the proposed Algorithm 1, Algorithm 2, Algorithm 3, and Algorithm 4.
    
    \item \emph{Fixed tasks offloading (FTO):}  Each device offloads $\beta_j^k=0.5 \alpha_j$ to its associated UAV. At the same time, sub-band assignment, power control, UAV deployment, and UAV task offloading problems are solved via the proposed Algorithm 1, Algorithm 2, Algorithm 3, and Algorithm 4. 
    
    \item \emph{Random sub-band assignment (RSA):} The available sub-bands in each UAV are randomly assigned to its associated devices, which offload their computation tasks to the UAV to perform remote computing. Device task offloading, power control, UAV deployment, and UAV task offloading problems are solved via Algorithm 2, Algorithm 3, and Algorithm 4.
    
    \item \emph{Fixed power Allocation (FPA):} Each device uses $50 \%$ of its maximum available power (i.e., $P_j^{k,b} = 0.5 P_j^{\mathbf{max}}$) in order to offload its computation task to UAVs to perform remote computing, while Algorithm 1, Algorithm 3, and Algorithm 4 are used to solve device task offloading, UAV deployment, and UAV task offloading problems.    
\end{itemize} 

 Furthermore, to evaluate the optimality gap of the proposed algorithm, we compare the performance of the proposed solution with the \emph{Optimal} scheme, where the sub-band assignment problem is solved by using the exhaustive search scheme, which can achieve the optimal solution. In contrast, the device task offloading, power control, UAV deployment, and UAV task offloading problems are solved via Algorithm 2, Algorithm 3, and Algorithm 4.
   \begin{figure}[t!]
   \centering
    \captionsetup{justification = centering}
    \includegraphics[width=0.5\linewidth]{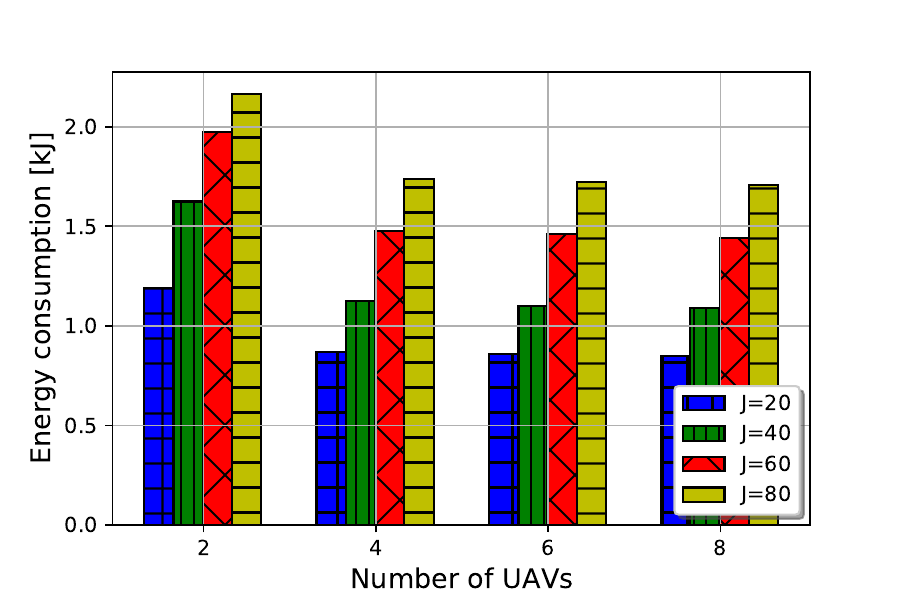}
    \caption{Energy consumption as a function of the number of UAVs for $J= 20, 40, 60$ and $80$.}
    \label{ene}
    \end{figure}

  \begin{figure}[t!]
   \centering
   \captionsetup{justification = centering}
   \includegraphics[width=0.5\linewidth]{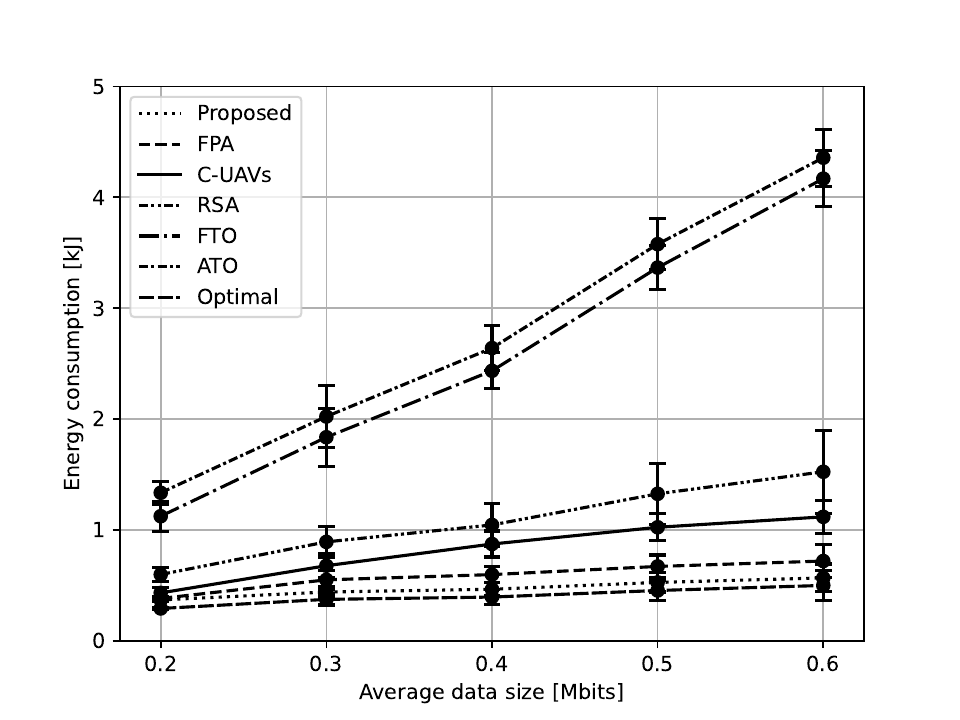}
   \caption{Energy consumption for $J=60$ devices under different average data sizes.}
   \label{datasize}
   \end{figure}
   
  \begin{figure}[t]
   \centering
   \captionsetup{justification = centering}
   \includegraphics[width=0.5\linewidth]{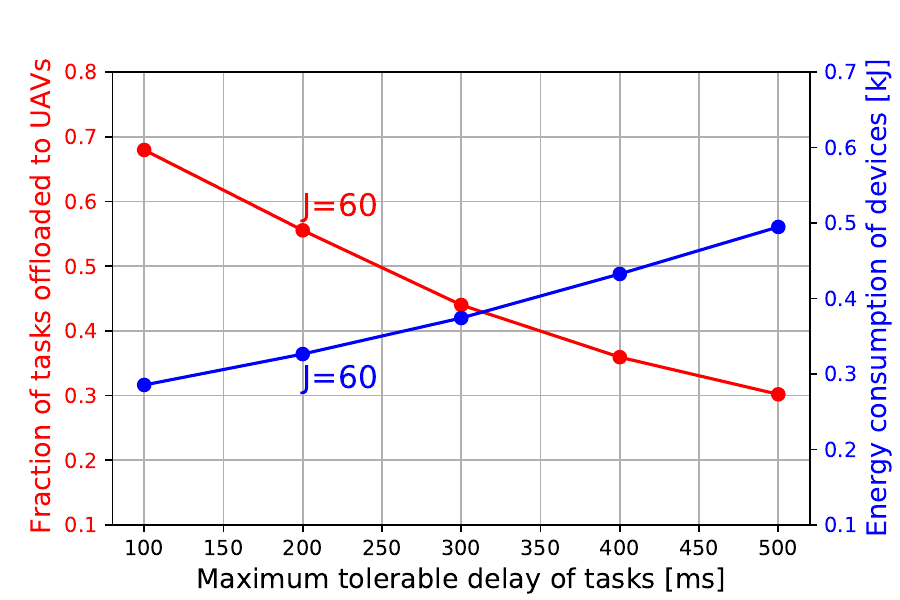}
   \caption{Fraction of tasks offloaded and energy consumption vs. maximum tolerable delay [ms].}
   \label{delaynocollab}
   \end{figure}
   
   \begin{figure}[t]
\centering
 \captionsetup{justification = centering}
 \includegraphics[width=0.5\linewidth]{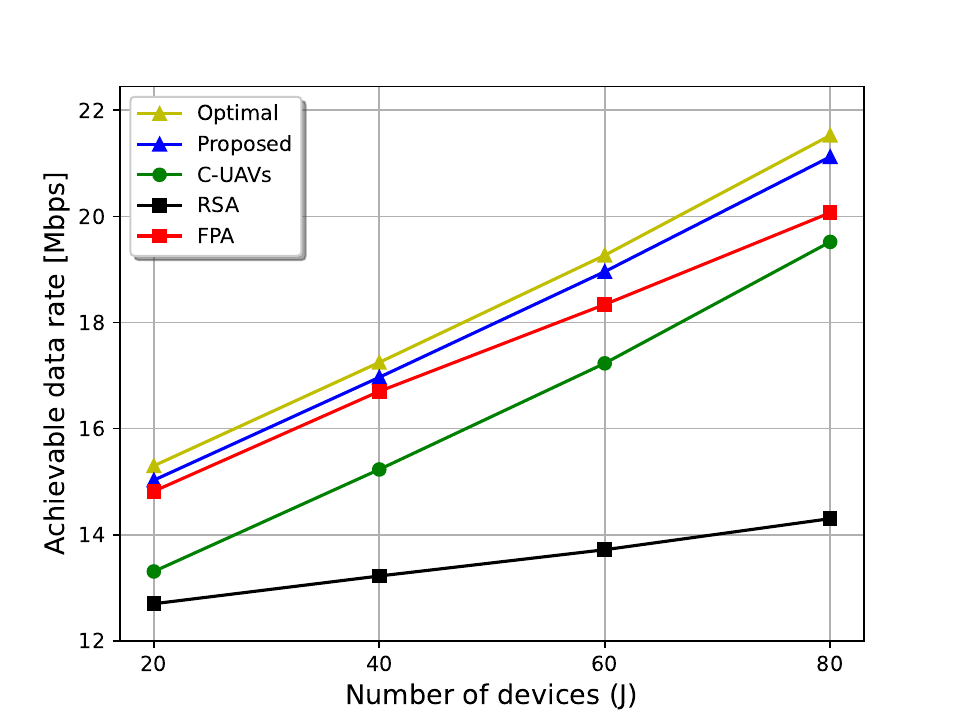}
 \caption{Comparison of the achievable data rate of devices.}
  \label{data}
 \end{figure} 

Fig.~\ref{devi} shows the energy consumption as a function of the number of devices in the network. The figure shows that the energy consumption under the ATO and FTO variants is significantly higher than under other variants of the proposed scheme. These results show that the most important optimization variable for minimizing energy consumption is the amount of data to be offloaded for computation. Additionally, compared to the C-UAVs and FPA variants, the energy consumption under the RSA variant is significantly higher than that of the proposed algorithm. As a result, we may conclude that compared to the deployment of UAVs and power control, sub-band assignment (i.e., $\boldsymbol{a}$) has a greater impact on energy consumption. In addition, as the number of devices in the network grows, the performance gap between our proposed solution and its variants which implies that optimization in all variables becomes increasingly important as the system size increases. Finally, the figure shows that the energy consumption under the proposed solution is nearly the same as that of the \emph{Optimal} scheme (i.e., the lower optimality gap).   
 \par 

Fig.~\ref{ene} shows the energy consumption as a function of the number of UAVs in the network. The figure shows that when deploying only 2 UAVs in the network, energy consumption is significantly higher than when there are 4, 6, and 8 UAVs in the network for all device counts, i.e., J = 20, 40, 60, and 80. However, it is interesting that the energy consumption under 4 UAVs, 6 UAVs, and 8 UAVs is nearly the same. Therefore, for the considered coverage area and device counts, deploying 6 UAVs and 8 UAVs will not give any benefit in terms of energy reduction, but will increase the hardware cost. In addition, when hovering energy for UAVs is taken into account, deploying 6 and 8 UAVs will even result in higher energy consumption than deploying 4 UAVs. Fig.~\ref{datasize} shows the energy consumption as a function of the average data size of the devices, together with the $95\%$ confidence intervals. The results show that the energy consumption increases approximately linearly with the average data size and confirm the importance of optimizing the fraction of data offloaded and of the sub-band assignment in minimizing the energy consumption (c.f., FTO, ATO, and RSA variants vs. proposed). Furthermore, the figure shows the lower optimality gap, proving the proposed solution's efficiency.

 \subsection{Impact of the Delay constraint} 
Fig.~\ref{delaynocollab} shows the average fraction of the devices' tasks that are offloaded as a function of the maximum allowable delay of tasks. The figure shows that the fraction of offloaded data decreases as the tasks' allowable delay increases. At the same time, the energy consumption of the devices increases. These results show that computation offloading in the considered system is essential for satisfying the tasks' delay constraints, but it leads to higher energy consumption than local computing.

\subsection{Data Rate Analysis}  
Fig.~\ref{data} shows the achievable data rate of the devices as a function of the number of devices, when using the proposed scheme and its variants. The results for the achievable data rate explain well the difference in terms of energy consumption among the variants of the proposed scheme. The data rate is lowest for the RSA variant, which explains why sub-band assignment is crucial for low energy consumption. We can also observe that the effect of not optimizing the UAVs' locations is significant, much bigger than that of not optimizing the transmit power allocation. Finally, the total data rate attained utilizing our proposed solution is the highest compared to alternative variants, and the proposed solution achieves nearly the same total data rate compared to that of the \emph{Optimal} scheme. These results explain why our proposed solution has the lowest energy consumption, according to (12) as shown in Fig.~\ref{devi}.

  \begin{figure}[t!]
   \centering
   \captionsetup{justification = centering}
   \includegraphics[width=0.5\linewidth]{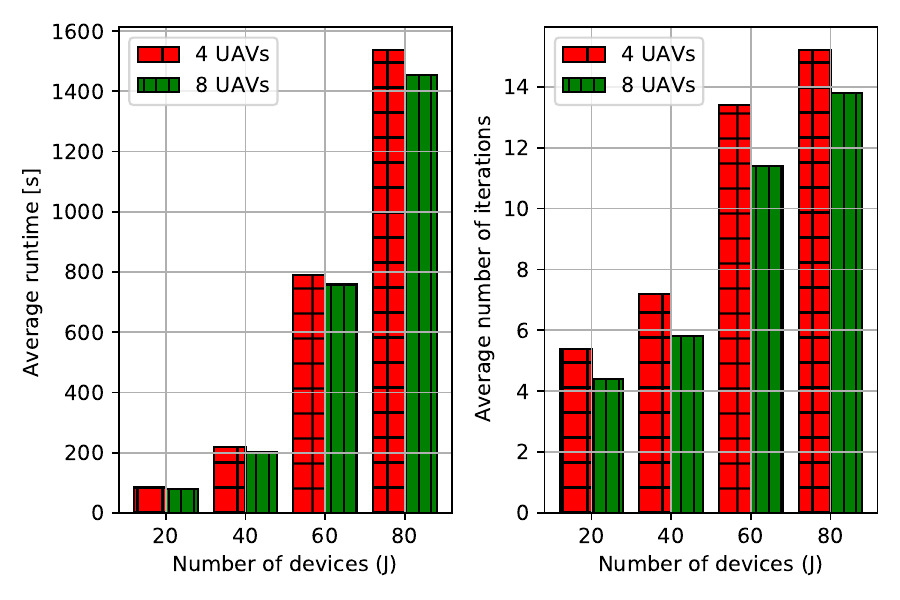}
   \caption{The average runtime $[s]$ and the average number of iterations for the convergence of the proposed algorithm.}
   \label{conv}
   \end{figure}

\subsection{Convergence of Proposed Algorithm}

Fig.~\ref{conv} shows the average runtime, and the average number of iterations to the convergence of the proposed solution as a function of the number of devices for $K = 4,$ and $8$. The figure shows that increasing the number of devices in the network results in the runtime growing considerably. However, the average number of iterations does not significantly increase. Additionally, it is interesting that deploying 4 UAVs requires more runtime and iterations to converge the proposed solution than deploying 8 UAVs. The reason is that deploying more UAVs will result in fewer associated devices at each UAV, which results in less burden to the UAV for decision making.

\section{Conclusions}
\label{con} 
In this paper, we considered THz-assisted MEC-enabled integrated SAG networks to provide computation services to wireless devices in remote areas. Then, we investigated the energy minimization problem by optimization tasks offloading decision, sub-bands assignment, power control, and UAVs deployment while guaranteeing the maximum tolerable delay of devices' computation tasks. Following, we showed that the formulated problem is a non-convex problem. Thus, to solve the problem, we decomposed the problem into four subproblems, namely, device task offloading decision problem, sub-band assignment and power control problem, UAV deployment problem, and UAV task offloading decision problem, respectively. Then, we solved the device task offloading decision problem by using the convex optimization technique, and a two-sided one-to-one matching game and CCP approach were deployed to address the sub-band assignment and power control problem. Moreover, we proposed SCA and BSUM to solve UAV deployment and UAV task offloading decision problems. Finally, we conducted comprehensive simulations to demonstrate the effectiveness of the proposed algorithm, and it was found that when compared to benchmark schemes, our proposed method significantly reduces the energy consumption of the UAVs and devices.

%\begin{appendices}  
%\section{Proof of Theorem 1}

%\end{appendices}

\ifCLASSOPTIONcaptionsoff
  \newpage
\fi
\bibliographystyle{IEEEtran}
\bibliography{ICC}
	
\end{document}